# Influence of Geometry and Tissue Damage on the Mechanical Behaviour of Skin Meshing

**Masoumeh Razaghi Pey Ghaleh [1,2,] *, Denis O'Mahoney[1,2], Kevin Moerman[3]**

[1] Mechanical and Industrial Engineering, School of Engineering, Atlantic Technological University, Galway, Ireland
[2] Centre for Mathematical Modelling and Intelligent Systems for Health and Environment (MISHE), School of Engineering, Atlantic Technological University, Sligo, Ireland
[3] Mechanical Engineering, School of Engineering, College of Science and Engineering, University of Galway, Ireland
* Correspondence: masoumeh.razaghipeyghaleh@research.atu.ie

## Abstract

Skin meshing increases graft coverage, yet the clinically achieved expansion often falls below the manufacturer's nominal meshing ratio. Previous elastic analyses have demonstrated that mesh topology and skin anisotropy strongly influence graft expansion; however, the effect of progressive damage at larger deformation remains unexplored. This study therefore investigates how progressive damage affects the expansion and mechanical response of skin meshes. The performance of auxetic mesh geometries is compared with conventional Zimmer mesher (ZM) geometry. Human back skin was modelled using a one-term Ogden ground matrix and a Gasser Ogden Holzapfel (GOH) collagen network, with separate scalar Kachanov-type continuum damage variables for each constituent. The auxetic geometries, including 1$^{st}$- and 2$^{nd}$-order alternative slits (AS1 and AS2), I-shaped (IS) and H-shaped (HS) re-entrant patterns, and ZM, were analysed with Langer's lines oriented either parallel or perpendicular to the loading direction. Hyperelastic and damage-based formulations were compared up to 25% strain, and the damage-based simulations were subsequently extended to 40% strain. ZM was further loaded to 155% strain; the area ratios attained by HS at 40% strain were matched by ZM at 142.1% and 124.7% strain for Langer's lines parallel and perpendicular to loading, respectively.

Up to a strain of 25%, damage had a negligible effect, and both damage-based and hyperelastic models produced closely comparable mechanical responses. Incorporating damage did not alter the expansion ranking of the geometries. At a strain of 40%, HS showed the greatest area ratios of 2.399 and 2.115 for Langer's lines parallel and perpendicular to the loading, respectively, corresponding to expansions 69.7% and 51.7% greater than ZM. This difference was associated with pronounced auxetic lateral expansion. In contrast, ZM exhibited negligible auxetic behaviour under stretch along Langer's lines, and slight lateral contraction in the perpendicular direction. However, it exhibited the lowest maximum damage, volume-weighted 90$^{th}$ percentile von Mises stress, and volume-averaged damage. To reach the HS area ratios, ZM required 142.1% and 124.7% strain in the parallel and perpendicular orientations, respectively; at these matched area ratios, HS showed lower volume-averaged damage than ZM in the parallel orientation but slightly higher values in the perpendicular orientation. Further ZM expansion increased its volume-averaged damage by 139.3% and 1,807.7% in the parallel and perpendicular orientations, respectively. These results indicate that HS achieves greater coverage than ZM for a given stretch, whereas its damage advantage depends on the orientation of Langer's lines.

## 1. Introduction

Severe burn injuries are commonly assessed by the percentage of total body surface area affected, with burns exceeding 30% requiring advanced clinical interventions such as skin grafting [1]. Among available treatments, skin meshing is widely adopted for its cost-effectiveness and ability to expand limited donor skin to cover large wound areas [1,2]. In this technique, controlled slit patterns are introduced into a graft, allowing it to stretch and open up to increase its area [1]. The effectiveness of this process is quantified by the meshing ratio, defined as the ratio of the expanded to the original skin area. However, a persistent discrepancy exists between the expansion ratios reported by manufacturers and those achieved clinically, so that additional donor skin must be harvested to cover the wound area [1,3]. Understanding the mechanical factors that limit graft expansion is therefore important for improving mesher design and achieving the expansion ratios demanded by the surgeon.

The accurate representation of skin meshing ratios is linked to the complex mechanical behaviour of skin [1]. Human skin consists of three primary layers: the epidermis, dermis, and hypodermis. While the epidermis and hypodermis are largely isotropic and contribute minimally to load-bearing, the dermis governs the mechanical response due to its dense network of collagen and elastin fibres [1,4]. The orientation and dispersion of collagen fibres, associated with Langer's lines, give rise to anisotropic behaviour, meaning that the mechanical response of skin depends strongly on the direction of loading [1,5–7]. Experimental studies have shown that skin is stiffer and exhibits higher tensile strength when stretched along these lines, while greater extensibility is observed in the perpendicular direction [1,5,6,8,9]. Consequently, both the magnitude and direction of deformation play a critical role in determining achievable meshing ratios [1].

Despite the importance of dermal anisotropy, many computational studies of skin meshing rely on simplified isotropic hyperelastic models, such as the Mooney–Rivlin, Yeoh, and Ogden formulations, which treat the dermis as the load-bearing component of skin meshing [1,3,6,10–15]. While these models can capture nonlinear elasticity, they fail to represent fibre-reinforced anisotropy and therefore cannot accurately predict large deformation behaviour during meshing. Recent studies have shown that structural constitutive models, particularly the Gasser Ogden Holzapfel (GOH) model, provide a more realistic framework by incorporating collagen fibre orientation and dispersion through a probabilistic distribution [1,3,7]. This model has demonstrated strong calibration agreement with experimental data while maintaining a relatively small number of parameters, making it suitable for simulating large deformations in skin meshing [1,7]. Furthermore, the results showed that expansion ratios, Poisson's ratios, reaction force, and stress were strongly dependent on fibre orientation and mesh pattern [3]. Using hyperelastic parameters calibrated for human back skin, a comparison of 1$^{st}$- and 2$^{nd}$-order alternative slits, I re-entrant slits (IS), and H re-entrant slits (HS) showed that HS exhibited the strongest auxetic response and, consequently, the greatest lateral and area expansion under 25% tensile stretch ($\lambda_{\text{load}} = 1.25$) [3].

However, isotropic and anisotropic hyperelastic formulations describe a purely elastic, recoverable response and therefore do not account for progressive, irreversible damage or loss of load-carrying capacity. Incorporating damage becomes increasingly critical at larger strains, where the skin meshing, particularly near the slits' tips, might rupture and inhibit further expansion of the skin meshes [1,7]. Li and Luo [16] modelled damage in human skin by extending the GOH formulation with Volokh-type functions [17,18], originally developed for arterial failure, for both the ground matrix and collagen fibres. Their damage-based GOH model, comprising nine material parameters, was fitted to uniaxial tensile data for human back skin, yielding a mean-normalised root-mean-square error of 8.02%. However, the fitted mean fibre orientation, $\gamma_{\text{fit}} = 0.0696°$, differed substantially from the histologically measured orientation of $\gamma_{\text{histology}} = \pm 41°$ (variability of approximately ±5°), suggesting that $\gamma$ could not be reliably identified from the mechanical data alone and was inconsistent with the known collagen architecture of human back skin [5,16].

To retain structural parameters consistent with human back skin, the present study developed an Ogden–GOH formulation with separate scalar Kachanov-type continuum damage variables [19,20] for the ground matrix and collagen fibre strain energy contributions. In continuum damage mechanics, these variables progressively reduce the material strain energy contributions, allowing stiffness reduction to be represented without explicitly modelling crack initiation or propagation. The ground matrix was described using a one-term Ogden formulation, whereas the GOH term represented the distributed collagen network. The elastic and damage parameters were determined through inverse finite element analysis of published uniaxial data on human back skin for loading parallel and perpendicular to Langer's lines. A staggered optimisation procedure was followed whereby the first step focused on the identification of the elastic parameters, using the pre-damage response. The next step then featured separate identification of the damage parameters of the ground matrix and collagen network. The data presented with an apparent fast and a more gradual damage process. The best match was obtained by associating the former with the collagen network and the latter with the ground matrix. Damage evolution was described using a log-normal cumulative distribution function (CDF), allowing damage to increase smoothly with strain. The resulting formulation used a total of nine fitted parameters (5 elastic parameters and 2 damage parameters each for the ground matrix and fibres, respectively) and reproduced the full experimental response with a sum of squared errors (SSE) of 0.003986, while maintaining structural parameters that are more consistent with values reported for human back skin.

The calibrated hyperelastic and damage-based Ogden–GOH parameter sets were implemented in finite element models of 1$^{\text{st}}$- and 2$^{\text{nd}}$-order alternative slits (AS1 and AS2), IS, HS, and the conventional Zimmer mesher (ZM), with a nominal expansion ratio of 1.5:1 [6]. Our previous study, featuring purely hyperelastic anisotropic formulations, showed that expansion generally increased as Langer's lines approached the loading direction, with HS exhibiting the greatest expansion, followed by IS, AS1, and AS2 [3]. In the present study, the auxetic and ZM geometries were analysed with Langer's lines oriented either parallel or perpendicular to the applied loading direction. The hyperelastic and damage-based formulations were first compared at $\lambda_{\text{load}} = 1.25$, representing the nominally elastic benchmark, after which the damage-based simulations were extended to $\lambda_{\text{load}} = 1.40$ to evaluate large-deformation damage. The aim was to determine whether the previously observed ranking of the auxetic geometries by area expansion persisted when progressive damage was incorporated, and to compare their deformation and mechanical responses with those of the ZM geometry. Performance was quantified using area ratio, apparent Poisson's ratio, transverse displacement, reaction force, volume-weighted 90th percentile von Mises stress, maximum damage, and volume-averaged damage. Furthermore, ZM loading was extended further to $\lambda_{\text{load}} = 2.55$ in order to reach expansion ratios comparable to HS at $\lambda_{\text{load}} = 1.40$, allowing the required stretch and the associated volume-averaged damage to be compared at matched expansion ratios.

The results show that HS retained the greatest expansion when damage was included, achieving area ratios 69.7% and 51.7% greater than ZM at $\lambda_{\text{load}} = 1.40$, while ZM required substantially larger stretch to reach comparable coverage.

Finally, the codes used for inverse finite element analysis and skin mesh geometry creation and finite element analysis have been made available as open source.

## 2. Materials and Methods

### 2.1. Finite element framework

All geometry generation, mesh construction, fibre orientation assignment, FEBio input-file preparation, and post-processing were performed using custom MATLAB R2026a routines developed with the open-source GIBBON toolbox [21]. The relevant codes have been made available as open source: https://github.com/MasoumehRazaghi90/damage-versus-elastic-modelling.

Nonlinear finite element analyses were performed using FEBio v4.0 [22]. Separate model-generation scripts were used for the Zimmer mesher geometry (ZM) and auxetic geometries to accommodate their distinct meshing, while maintaining consistent material definitions, fibre orientations, boundary conditions, and post-processing procedures.

The analyses comprised two simulation sets. Firstly, the hyperelastic Ogden–GOH formulation was compared with its damage-based counterpart at $\lambda_{\text{load}} = 1.25$, enabling direct comparison with the hyperelastic stretch range. Secondly, all damage-based simulations were extended to $\lambda_{\text{load}} = 1.40$, where unmeshed skin loaded parallel to the collagen fibres had entered the experimental damage range. Each geometry was analysed with Langer's lines parallel and perpendicular to loading to assess mechanically sustainable expansion and progression towards severe local damage.

Nodal displacements, reaction forces, and element-level principal stresses were exported from FEBio. The damage-based analyses additionally provided element-level strain energy density and scalar damage. These outputs were processed in MATLAB to calculate the area ratio, Poisson's ratio, maximum transverse displacements, axial reaction force, volume-weighted 90$^{\text{th}}$ percentile von Mises stress, maximum damage, and volume-averaged damage. These quantities characterised mesh expansion, mechanical response, and the severity of damage. This staged workflow enabled direct comparison of the hyperelastic and damage-based predictions at $\lambda_{\text{load}} = 1.25$ (elastic range), followed by the evaluation of expansion and mechanical integrity at $\lambda_{\text{load}} = 1.40$ (damage range), where progressive damage was quantified. Additional damage-based ZM simulations were extended to $\lambda_{\text{load}} = 2.55$ to reach area ratios comparable to those of HS, which showed the greatest expansion. The comparison examined the stretch required to achieve similar area expansion and the associated volume-averaged damage.

### 2.2. Finite element model construction

Five slit-based meshing geometries were generated: 1$^{\text{st}}$- and 2$^{\text{nd}}$-order alternative slits (AS1 and AS2), I-shaped and H-shaped re-entrant slits (IS and HS), and a conventional Zimmer pattern (ZM) (Figure 1), with the ZM having a nominal expansion ratio of 1.5:1 [3,6]. The geometric parameters defining the individual patterns are summarised in Table 1 and illustrated in Figure 1. All models had comparable in-plane dimensions of approximately 60 mm and a thickness of 3 mm, thereby reducing the influence of specimen size and thickness on comparisons among the different mesh configurations. The four auxetic patterns were constructed with a nominal cut width of 1 mm. In contrast, the ZM pattern used a 29 mm slit length with 3 mm inter-slit spacing; the slits were represented by duplicating coincident nodes along the prescribed cut lines and reassigning adjacent elements, thereby producing slits with zero initial width (0 mm cuts).

The auxetic geometries were constructed from closed boundary curves defining one-quarter of each unit cell. These boundary curves were resampled to obtain a uniform desired node spacing, and surfaces were subsequently discretised using constrained Delaunay triangulation with the *regionTriMesh* function in GIBBON [21]. Triangulations were used to conform to the irregular slit and re-entrant boundaries while permitting local mesh refinement. Local mesh refinement was applied near slit tips and re-entrant corners using *subTriLocal*, where elevated deformation and stress gradients were anticipated. Each meshed quarter cell was mirrored in the two in-plane directions to generate a complete unit cell. The resulting unit cells were then replicated and translated to form the full periodic patterns, after which coincident vertices at reflection boundaries and interfaces between adjacent cells were merged to maintain mesh continuity. The planar meshes were subsequently extruded to a thickness of 3 mm, generating three-dimensional models composed of six-noded pentahedral elements [3].

**Table 1.** Parameters for the Auxetic and conventional skin meshing pattern geometries in (mm).

| **Auxetic pattern** | **C1** | **C2** | **D1** | **D2** | **T** | | | |
|---|---|---|---|---|---|---|---|---|
| **AS 1$^{st}$-order** | 3.8 | 3.8 | 22.04 | 22.04 | 1 | | | |
| **Auxetic pattern** | **L** | **N** | **S2** | **S3** | **T** | | | |
| **IS** | 9.5 | 7.5 | 7.5 | 1 | 1 | | | |
| **Auxetic pattern** | **H1** | **H2** | **S1** | **S2** | **S3** | **S4** | **N** | **T** |
| **AS 2$^{nd}$ -order** | 15.36 | 7.68 | 0.66 | 0.46 | 1.42 | 1.82 | 8.6 | 1 |
| **Auxetic pattern** | **L1** | **L2** | **S1** | **S2** | **S5** | **T** | | |
| **HS** | 7.605 | 17.37 | 1.08 | 2.25 | 0.99 | 1 | | |
| **Conventional pattern** | **L1** | **L2** | | | | | | |
| **ZM** | 29 | 3 | | | | | | |

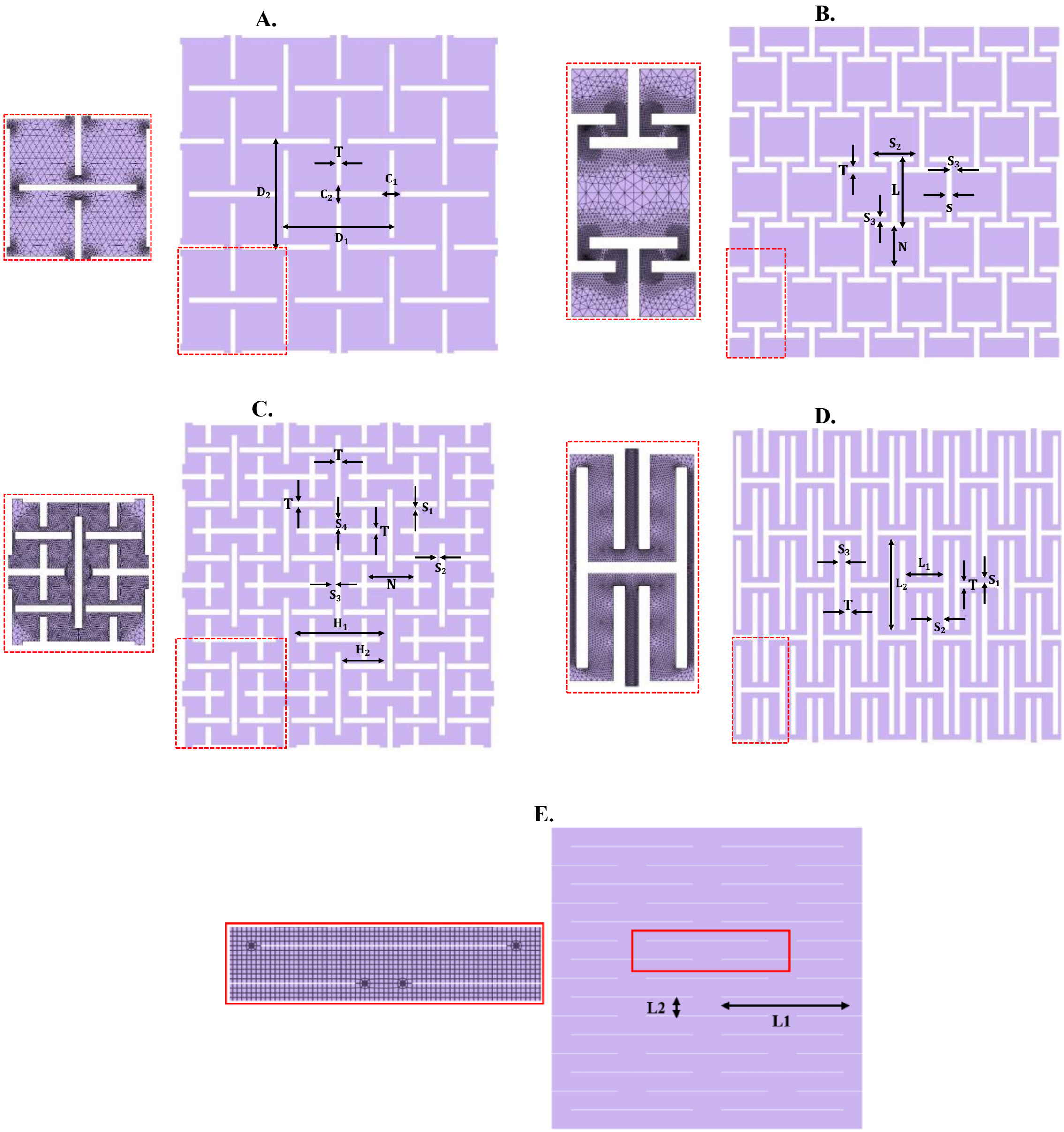

**Figure 1**. Parametric finite element models for auxetic skin graft patterns: A) Alternative slit 1st-order (AS1), B) I re-entrant slit (IS), C) Alternative slit 2nd-order (AS2), D) H re-entrant slit (HS), E) Zimmer pattern with nominal expansion ratio of 1.5:1.

The ZM geometry was generated separately because its staggered slit arrangement allowed for a simpler meshing procedure. A structured quadrilateral mesh was created with 29-mm slits and 3-mm inter-slit spacing, arranged in alternating staggered rows. Mesh density around the slit tips was increased by six successive levels of quadrilateral refinement, each splitting an element into four smaller elements. The slits were then formed by duplicating coincident nodes along the prescribed cut paths and reassigning the connectivity of adjacent elements, thereby representing zero-initial-aperture discontinuities rather than finite-width voids. The completed planar mesh was then extruded to a thickness of 3 mm to generate eight-noded hexahedral elements.

Comparable specimen dimensions and a common thickness of 3 mm were maintained across the models. The model thickness was selected to match the mean full thickness of human back skin, reported as 3.04 ± 0.63 mm [3,23]. All meshed geometries converged with approximately 10,000 elements for the auxetic geometries and approximately 21,000 elements for ZM.

## 2.3. Constitutive modelling

The mechanical response of skin was represented using an anisotropic Ogden–GOH solid mixture in both hyperelastic and damage-based forms. The non-collagenous ground matrix was described by a one-term Ogden model, whereas the distributed collagen network was represented by two symmetric GOH fibre families. The Ogden model was selected because its exponent captures the nonlinear response of the ground matrix, which a Neo-Hookean matrix could not reproduce during calibration (Section 2.4).

The damage-based formulation retained the same hyperelastic Ogden–GOH model but introduced separate scalar Kachanov-type damage variables [19,20] for the strain energy density functions of the ground matrix and collagen fibres. This separation allowed the two constituents to sustain damage at different rates. The fibre-associated process was formulated to evolve more rapidly after collagen recruitment (parallel to Langer's lines), whereas the matrix-associated process was formulated to represent a more gradual reduction in the load-bearing contribution of the ground matrix (perpendicular to Langer's lines). Although the model is phenomenological, this separation was necessary, as a single damage law could not reproduce the distinct damage responses observed in the two loading directions (Section 2.4).

Damage evolution and the staggered calibration are described in Sections 2.3.2 and 2.4, respectively.

### *2.3.1. Hyperelastic Ogden-GOH model*

The hyperelastic strain energy density function can be written as:

$$\Psi_{\text{hyperelastic}} = \Psi_{\text{Ogden}} + \Psi_{\text{GOH}} \tag{1}$$

where $\Psi_{\text{Ogden}}$ denotes the ground-matrix contribution and $\Psi_{\text{GOH}}$ denotes the contribution of the collagen fibre families.

The ground matrix is represented by the one-term Ogden model [24]

$$\Psi_{\text{Ogden}} = \frac{c_1}{m_1^2}\left(\tilde{\lambda}_1^{m_1} + \tilde{\lambda}_2^{m_1} + \tilde{\lambda}_3^{m_1} - 3\right) + \frac{K_{\text{Ogden}}}{2}(\ln J)^2 \tag{2}$$

where $c_1$ is a stiffness parameter, $m_1$ controls the nonlinear matrix response, and $\frac{K_{\text{Ogden}}}{2}(\ln J)^2$ is the volumetric penalty, in which the bulk modulus $K_{\text{Ogden}}$ enforces near-incompressibility of the ground matrix. The deviatoric principal stretches are given by

$$\tilde{\lambda}_i = J^{-1/3}\lambda_i \tag{3}$$

where $\lambda_i$ are the principal stretches and $J = \lambda_1\lambda_2\lambda_3$ is the Jacobian, representing the volume ratio.

The collagen contribution is described by the GOH model [25]

$$\Psi_{\mathrm{GOH}} = \frac{c}{2}(\tilde{I}_1 - 3) + \frac{k_1}{2k_2}\sum_{i=1}^{2}\left(\exp\left(k_2\langle\tilde{E}_i\rangle^2\right) - 1\right) + \frac{K_{\mathrm{GOH}}}{2}\left(\frac{J^2-1}{2} - \ln J\right) \quad (4)$$

where $c$ and $K_{\mathrm{GOH}}$ are the shear and bulk moduli of the isotropic ground-matrix part of the GOH model, $k_1$ controls fibre stiffness, and $k_2$ governs nonlinear fibre stiffening. Because the ground matrix is already represented by the Ogden model, $c$ was assigned a negligible value of 0.0001 MPa. Both bulk moduli were defined using the same multiplier, $K_{\mathrm{Ogden}} = 100c_1$ and $K_{\mathrm{GOH}} = 100c$, in the skin-meshing simulations, so that $K_{\mathrm{GOH}}$ is also negligible. The GOH model therefore contributes only the collagen fibre response, while the volumetric response, and hence near-incompressibility, is governed by $K_{\mathrm{Ogden}}$. The quantity $\tilde{I}_1 = tr\tilde{\mathbf{C}}$ is the first invariant of the deviatoric right Cauchy–Green tensor $\tilde{\mathbf{C}}$. The Macaulay brackets ensure that the collagen contribution is active only when the dispersed fibre strain measure is positive.

The dispersed fibre strain measure is defined as

$$\tilde{E}_i = \kappa(\tilde{I}_1 - 3) + (1 - 3\kappa)(\tilde{I}_{4i} - 1) \quad (5)$$

where $\tilde{I}_{4i} = \boldsymbol{a}_{\mathrm{i}} \cdot \tilde{\mathbf{C}}\boldsymbol{a}_{\mathrm{i}}$ is the squared stretch along the mean direction of the $i^{th}$ fibre family, and $\kappa \in \left[0, \frac{1}{3}\right]$ describes fibre dispersion. The two mean fibre direction vectors were defined symmetrically within the local material plane as $\boldsymbol{a}_{1,2} = \cos\gamma\ \mathbf{e_1} \pm \sin\gamma\mathbf{e_2}$ where $\mathbf{e_1}$ is aligned with Langer's lines and $\mathbf{e_2}$ is the orthogonal in-plane direction, so that the fibres form angles of $\pm\gamma$ with respect to $\mathbf{e_1}$. The limiting value $\kappa = 0$ corresponds to perfectly aligned fibres, and $\kappa = \frac{1}{3}$ represents an isotropic orientation distribution, whereas $0 < \kappa < \frac{1}{3}$ defines the dispersed orientation of collagen fibres.

### *2.3.2. Damage-based Ogden-GOH model*

Progressive damage is represented by applying independent scalar Kachanov-type damage variables [19,20] to the Ogden ground matrix and GOH collagen fibre contributions, whereby each strain energy density is multiplied by the integrity factor $(1 - D)$, following the finite-strain formulation of Simo [20]. The strain energy density of damaged material is then expressed as

$$\Psi_{\mathrm{damage}} = \left(1 - D_{\mathrm{Ogden}}\right)\Psi_{\mathrm{Ogden}} + (1 - D_{\mathrm{GOH}})\Psi_{\mathrm{GOH}} \quad (6)$$

where $\Psi_{\mathrm{Ogden}}$ and $\Psi_{\mathrm{GOH}}$ are the strain energy densities defined in equations (2) and (4), while $D_{\mathrm{Ogden}}$ and $D_{\mathrm{GOH}}$ quantify damage of the ground matrix and collagen fibre contributions, respectively.

For each constituent $r \in \{\mathrm{Ogden, GOH}\}$, the damage variable satisfies

$$0 \leq D_{\mathrm{r}} \leq D_{\mathrm{r,max}} \leq 1, \quad (7)$$

An undamaged constituent is represented by $D_\mathrm{r} = 0$. As $D_\mathrm{r}$ increases, the stress and strain energy are reduced by the factor $(1 - D_\mathrm{r})$. Here $D_\mathrm{r,max} = 1$ was used for both constituents; hence, $D_\mathrm{r} = 1$ represents complete damage.

Damage evolution is defined using a log-normal cumulative distribution function (CDF), following the FEBio implementation [22]:

$$\mathrm{CDF}_r(\eta) = \frac{1}{2}\left[1 + \mathrm{erf}\left(\frac{\ln(\eta/\mu_r)}{\sqrt{2}\sigma_r}\right)\right], \qquad \eta > 0 \tag{8}$$

with $\mathrm{CDF}_r = 0$ for $\eta \leq 0$. The damage-driving variable $\eta$ is the largest positive maximum principal Green–Lagrange strain attained at each material point during the loading history:

$$\eta(t) = \max_{0 \leq \tau \leq t} \langle E^\mathrm{L}_\mathrm{max}(\tau) \rangle \tag{9}$$

where $E^\mathrm{L}_\mathrm{max}$ is the current maximum principal Green–Lagrange strain and $\langle x \rangle = \max(\mathrm{x}, 0)$. Any damage developed is irreversible and permanent. The parameter $\mu_r$ is the median of the log-normal distribution and thus defines the maximum principal Green–Lagrange strain at which half of the maximum damage is reached, whereas $\sigma_r$ is the standard deviation of $\ln \eta$ and controls the spread of damage around $\mu_r$. A smaller $\sigma_r$ produces a more abrupt increase in damage, while a larger value distributes damage over a broader strain range.

The scalar damage variable is then calculated as

$$D_\mathrm{r}(\eta) = D_\mathrm{r,max}\,\mathrm{CDF}_r(\eta) \tag{10}$$

where $D_\mathrm{r,max}$ specifies the maximum permitted damage. Thus, the CDF governs how damage develops with increasing deformation, while $D_\mathrm{r,max}$ defines its limiting magnitude. By assigning different $\mu_r$ and $\sigma_r$ values to the Ogden and GOH constituents, the ground matrix and fibre-associated damage processes evolved independently.

## 2.4. Constitutive parameter calibration

Published uniaxial tensile data for human back skin were obtained from Figure 6 (case J) of Li and Luo [16]. Separate stress–stretch curves for loading parallel and perpendicular to Langer's lines extended from the initial nonlinear stiffening response into the large-strain damage regime.

Inverse FEA calibration was performed using the GIBBON toolbox [21] in MATLAB coupled with FEBio [22]. A single hexahedral element was subjected to uniaxial loading conditions as per the experimental stretch-stress history. The parallel and perpendicular Langer's-line orientations were represented by changing the orientation of the local material axes relative to the fixed loading direction while retaining the same specimen geometry and boundary conditions.

The fibre orientation was prescribed as $\gamma = \pm 41°$ from histological measurements of human back skin [5] rather than calibrated, and the GOH shear modulus $c$ and the bulk moduli (Section 2.3.1) were also held fixed.

The remaining constitutive parameters (see Table 2) were calibrated by minimising the sum of squared differences between the experimental and simulated stress–stretch curves using the Levenberg–Marquardt nonlinear least-squares algorithm [26]. A staggered calibration strategy was used, in which the hyperelastic parameters were first identified from the portion of each curve preceding the first signs of softening, considered undamaged, and then the damage parameters governing progressive damage were calibrated.

In the first hyperelastic calibration stage, only the data preceding the first signs of softening were used, namely up to $\lambda = 1.256$ for loading parallel and $\lambda = 1.477$ for loading perpendicular to Langer's lines. The ground matrix was represented by a one-term Ogden model with the exponent fixed at $m_1 = 2$, yielding a Neo-Hookean-type matrix response within the Neo-GOH formulation. The parameters $c_1$, $k_1$, $k_2$, and $\kappa$ were calibrated simultaneously against the parallel and perpendicular responses, while $m_1$, $\gamma$, and $c$ were held constant. The resulting Neo-GOH fits are shown in Figure 2(A).

The calibrated Neo-GOH parameters were then held fixed, and the analysis was extended over the full experimental stretch ranges of $\lambda = 1.441$ for parallel loading and $\lambda = 1.860$ for perpendicular loading. A log-normal damage law was applied to the combined Neo-GOH strain energy contribution, with only the location $(\mu)$ and dispersion $(\sigma)$ parameters calibrated at this stage. The maximum damage parameter was fixed at $D_{\max} = 1$, permitting complete continuum damage. The resulting unified-damage Neo-GOH response and strain-energy evolution are shown in Figure 2(B). The final sum of squared errors was 0.8204, indicating that the unified-damage formulation did not adequately capture the experimentally observed damage in both loading directions.

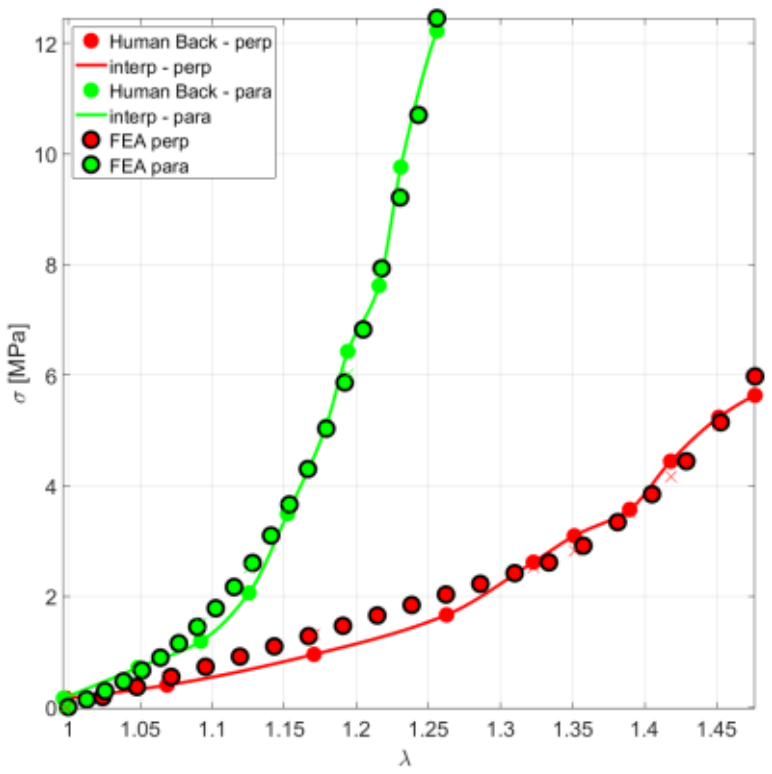

A.

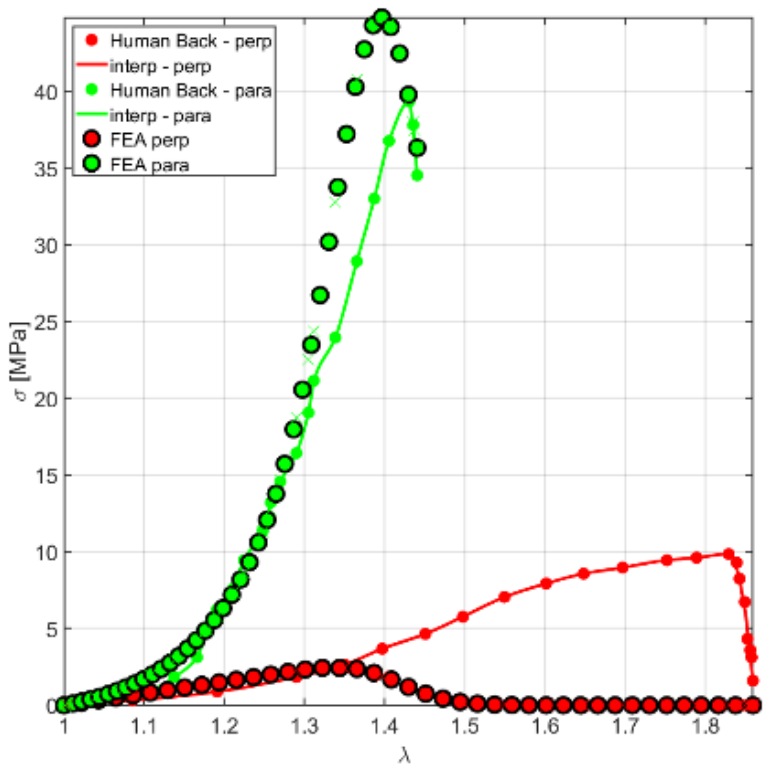

B.

**Figure 2**. Staggered calibration of the hyperelastic and damage-based constitutive formulations against published uniaxial tensile data for human back skin loaded parallel and perpendicular to Langer's lines [16]. (A) Pre-damage calibration of the Neo-GOH formulation, in which the one-term Ogden exponent was fixed at $m_1 = 2$. (B) Calibration of a unified damage law applied to the combined Neo-GOH over the complete experimental deformation range.

The hyperelastic calibration was therefore repeated using the general one-term Ogden formulation, with the exponent $m_1$ treated as an additional optimisation variable. The parameters $c_1, m_1, k_1, k_2$, and $\kappa$ were calibrated over the same pre-damage ranges, while $\gamma$ and $c$ were held constant. The fitted exponent was $m_1 = 6.943$, and the hyperelastic sum of squared errors decreased from 3.2004 for the Neo-GOH model to 0.00608 for the general Ogden–GOH model. The corresponding fit is shown in Figure 3(A).

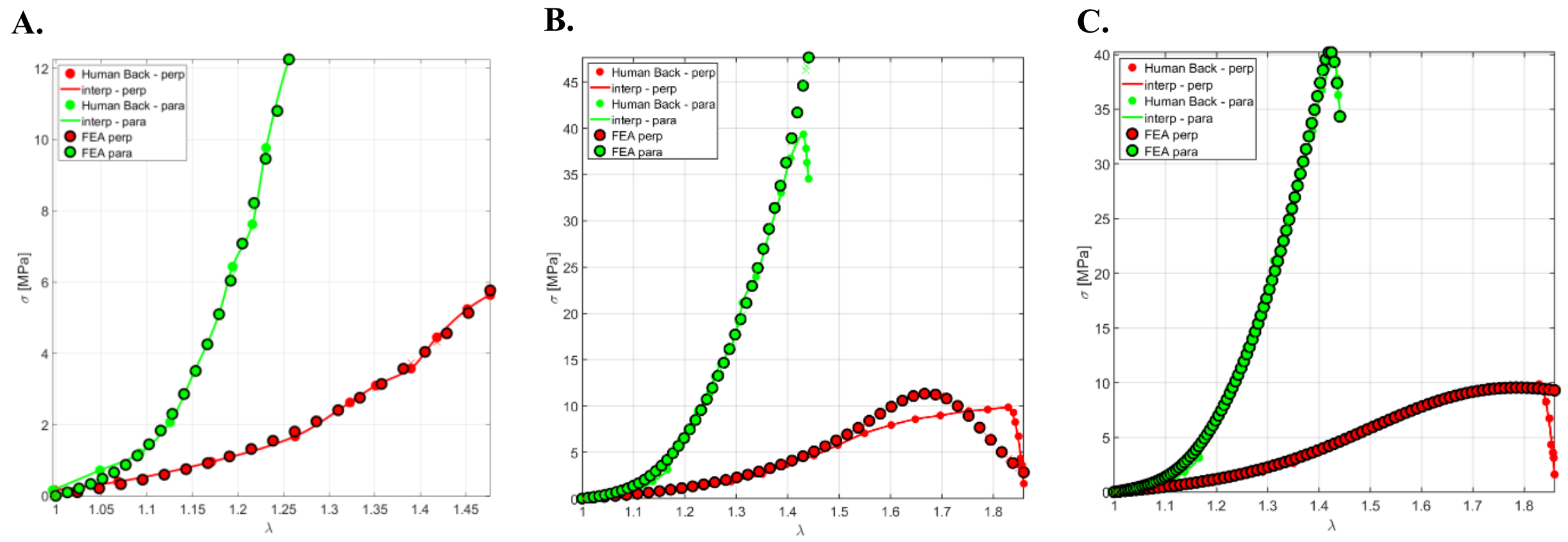

**Figure 3**. Staggered calibration of the hyperelastic and damage-based constitutive formulations against published uniaxial tensile data for human back skin loaded parallel and perpendicular to Langer's lines [16]. (A) Pre-damage calibration of the general Ogden–GOH formulation, in which $m_1$ was included as an optimisation variable. (B) Calibration of a unified damage law applied to the combined Ogden–GOH strain energy contribution. (C) Final staggered calibration stage, in which the hyperelastic and GOH damage parameters were fixed, and only the Ogden damage parameters were calibrated from the perpendicular loading response.

After calibrating the Ogden–GOH hyperelastic parameters, a single log-normal damage law was applied to the combined matrix and fibre strain energy functions. Only the location ($\mu$) and dispersion ($\sigma$) parameters were calibrated over the full parallel and perpendicular datasets, while $D_{\max}$ remained prescribed. The unified-damage Ogden–GOH model reduced the sum of squared errors to 0.02392, as shown in Figure 3(B). Although this represented a substantial improvement over the unified-damage Neo-GOH formulation, a single common damage law was still unable to accurately reproduce the distinct damage responses observed in the two loading directions.

The unified damage formulation was therefore replaced by separate constituent-specific damage laws for the GOH collagen network and the Ogden ground matrix, while the previously calibrated hyperelastic parameters were retained. First, the GOH damage parameters $\mu_{\text{GOH}}$ and $\sigma_{\text{GOH}}$ were calibrated to the more rapid damage observed under parallel loading. These parameters were then fixed, and the Ogden matrix damage parameters $\mu_{\text{Ogden}}$ and $\sigma_{\text{Ogden}}$ were calibrated to the more gradual perpendicular loading response. The corresponding limiting damage values were prescribed as $D_{\text{GOH,max}} = D_{\text{Ogden,max}} = 1$. This sequential procedure was used solely for parameter identification and does not imply that fibre damage occurs only under parallel loading or that matrix damage occurs only under perpendicular loading. For this final stage, the perpendicular data were used up to $\lambda = 1.82$; the remaining points at nearly constant stretch, corresponding to terminal specimen rupture, were excluded.

The resulting fit is shown in Figure 3(C). The complete constitutive formulation therefore comprises nine calibrated parameters: five shared hyperelastic parameters $(c_1, m_1, k_1, k_2, \kappa)$ and two constituent-specific damage-parameter pairs $(\mu_{\text{GOH}}, \sigma_{\text{GOH}}, \mu_{\text{Ogden}}, \sigma_{\text{Ogden}})$.

The final two-process damage formulation yielded a combined sum of squared errors of 0.003986, corresponding to reductions of 83.33% relative to the unified-damage Ogden–GOH model and 99.51% relative to the unified-damage Neo-GOH model. The prescribed and calibrated parameters for the final constitutive formulations are summarised in Table 2. The resulting Ogden–GOH hyperelastic parameters and constituent-specific damage parameters were subsequently used in the skin meshing simulations.

**Table 2.** Calibrated parameters of human back skin using a staggered optimisation

| Model | $c_1$[MPa] | $m_1$[MPa] | $k_1$ | $k_2$ | $\kappa$ | $\mu$ | $\sigma$ | | |
|---|---|---|---|---|---|---|---|---|---|
| **damage NEO-GOH** | 5.085 | 2 | 24.4 | 33.86 | 0.24 | 0.479 | 0.136 | | |
| **Model** | $c_1$**[MPa]** | $m_1$**[MPa]** | $k_1$ | $k_2$ | $\kappa$ | $\mu_{\text{Ogden}}$ | $\sigma_{\text{Ogden}}$ | $\mu_{\text{GOH}}$ | $\sigma_{\text{GOH}}$ |
| **damage Ogden-GOH** | 2.727 | 6.943 | 21.01 | 0.694 | 0.145 | 1.03 | 0.378 | 0.551 | 0.0472 |

## 2.5. Boundary conditions and finite element analysis

Following mesh generation, the calibrated Ogden–GOH hyperelastic formulation and its damage-based counterpart, described in Sections 2.3 and 2.4, were assigned to the elements through the local material axes. The four auxetic geometries (AS1, AS2, IS, and HS) and the Zimmer mesher geometry (ZM) were analysed with Langer's lines oriented either perpendicular or parallel to the applied tensile loading direction. Material anisotropy and its influence on the evolution of deformation and damage were thereby evaluated. Langer's line orientations of 0°and 90°were used to represent the perpendicular and parallel configurations, respectively.

For all geometries, tensile loading was applied along the global Y-direction. The ZM geometry was rotated by 90° about the global Z-axis before loading to maintain the same loading direction and orientation convention as the auxetic geometries. The applied loading stretch was defined as

$$\lambda_{\text{load}} = \frac{h}{h_0} = 1 + \frac{u_y}{h_0} \tag{11}$$

where $h_0$ and $h$ are the initial and current specimen dimensions in the loading direction, respectively, and $u_y$ is the prescribed displacement in the global Y-direction.

The same boundary conditions were applied to each model configuration (see Figure 4). The upper loading boundary was prescribed a Y-displacement and constrained against out-of-plane motion in Z, while remaining free in the transverse X-direction. The lower support boundary was constrained in Y and Z but remained free in X, permitting lateral expansion or contraction. A single reference node on the lower boundary was additionally constrained in X to remove rigid-body translation without suppressing transverse deformation.

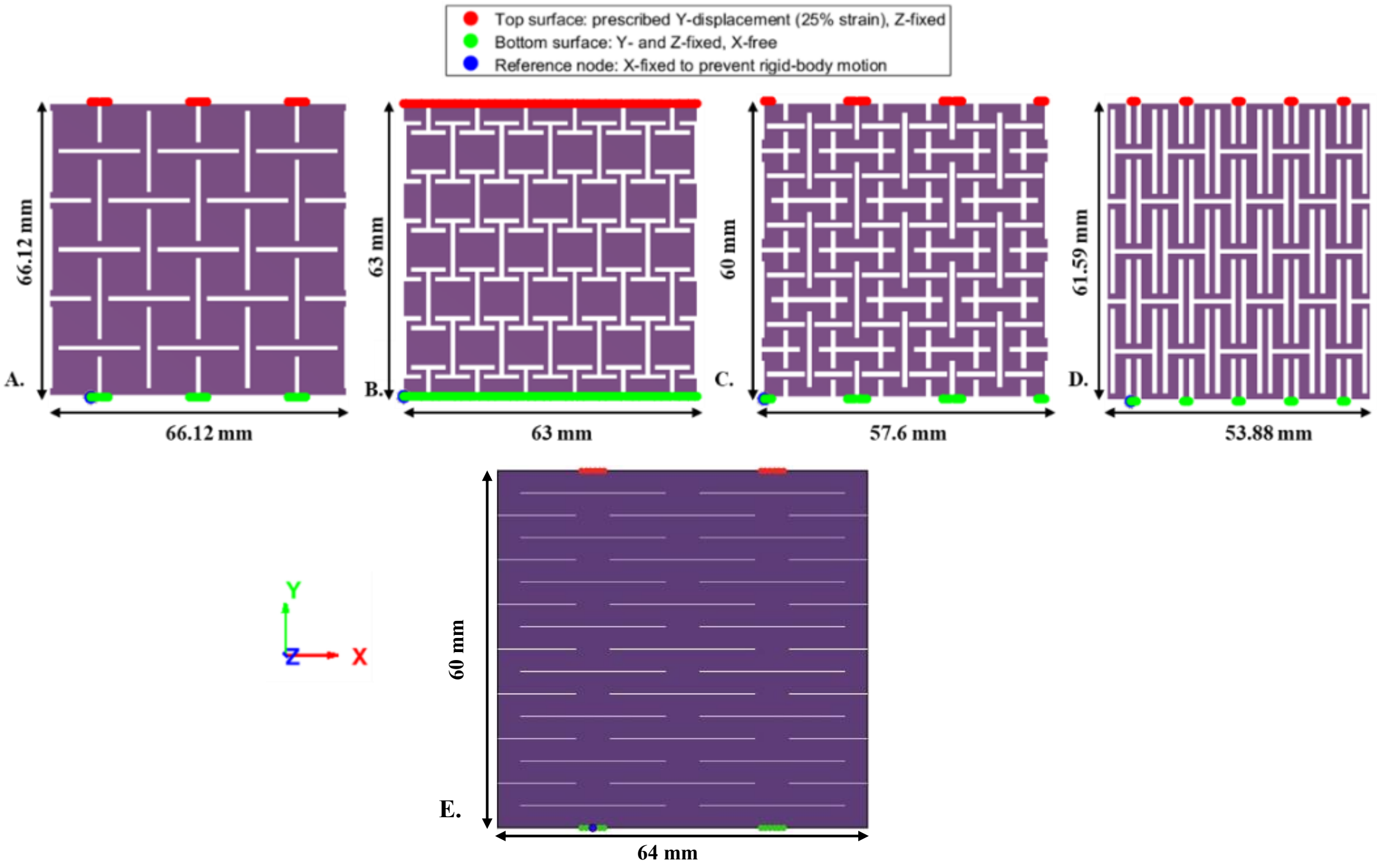

**Figure 4.** Boundary conditions applied to the five FEA models used in the uniaxial tensile simulations: (A) AS 1st order, (B) IS, (C) AS 2nd order, (D) HS, and (E) ZM. Red markers indicate the top surface with prescribed Y-displacement, green markers indicate the bottom surface constrained in Y and Z, and the blue marker indicates the reference node fixed in X to prevent rigid-body motion. The prescribed displacement corresponds to $\lambda_{\text{load}}$ up to 1.40 (2.55 for the extended ZM simulations)

## 2.6. Post-processing and model comparison

The FEBio outputs described in Section 2.1 were processed in MATLAB using the same framework for all geometries, with the global Y- and X-directions defined as the loading and transverse directions, respectively. The measures are defined below; transverse-displacement and damage fields were additionally visualised to examine their spatial distributions.

For each state, the transverse width $w$ and loading-direction height $h$ were obtained from the maximum and minimum of the deformed nodal coordinates. With $w_0$ and $h_0$ denoting the corresponding initial dimensions,

$$\lambda_x = \frac{w}{w_0}, \qquad \lambda_y = \frac{h}{h_0} \tag{12}$$

where $\lambda_y = \lambda_{\text{load}}$ and $\lambda_x = \lambda_{\text{transverse}}$. The in-plane area ratio was calculated as

$$R_A = \frac{A}{A_0} = \frac{wh}{w_0 h_0} = \lambda_x \lambda_y \tag{13}$$

where $A_0$ and $A$ are the initial and deformed planar areas, respectively. Thus, $R_A = 1$ represents no change in projected graft coverage, whereas $R_A > 1$ indicates expansion.

The effective Poisson's ratio was calculated from the transverse and loading stretches as

$$\upsilon = -\frac{\ln \lambda_{\mathrm{x}}}{\ln \lambda_{\mathrm{y}}} \tag{14}$$

Negative Poisson's ratios indicate transverse expansion during axial extension and are consistent with an auxetic structural response.

Mechanical resistance to expansion was quantified by the magnitude of the resultant reaction force in the global Y-direction. Local stress was characterised using the von Mises stress. To reduce sensitivity to local stress concentrations, overall model stress comparisons were based on the reference volume-weighted 90$^{\mathrm{th}}$ percentile of the von Mises stress. This was defined as

$$\sigma_{\mathrm{VM}}^{\mathrm{P90}} = \min\left\{s : \frac{\sum_{i:\sigma_{\mathrm{VM},i} \leq s} V_i^0}{\sum_i V_i^0} \geq 0.90\right\} \tag{15}$$

where $V_i^0$ is the reference volume of the element $i$, and $\sigma_{\mathrm{VM},i}$ is its von Mises stress. Thus, $\sigma_{\mathrm{VM}}^{\mathrm{P90}}$ is the lowest stress value $s$ at or below which 90% of the total reference volume lies.

For the damage-based simulations, the severity of local damage was characterised by the maximum element damage,

$$D_{\mathrm{peak}} = \max_i(D_i) \tag{16}$$

The overall level of damage was quantified using the reference-volume-weighted mean damage, referred to here as volume-averaged damage,

$$f_d = \frac{\sum_{i=1}^{N_e} D_i V_i^0}{\sum_{i=1}^{N_e} V_i^0} \tag{17}$$

where $D_i$ is the scalar damage variable of element $i$, $V_i^0$ is its reference volume, and $N_e$ is the total number of elements. All elements were included in this calculation, including those with zero damage. Whereas $D_{\mathrm{peak}}$ identifies the greatest local damage, $f_d$ characterises the average damage across the meshed skin, accounting for both its severity and spatial extent. Volume-averaged damage was plotted against area ratio to assess the damage associated with graft expansion.

Area ratio, apparent Poisson's ratio, axial reaction force, and reference-volume-weighted P90 von Mises stress were compared between the hyperelastic and damage-based formulations over the common loading range up to $\lambda_{\mathrm{load}} = 1.25$. The damage-based response was further evaluated up to $\lambda_{\mathrm{load}} = 1.40$, using maximum damage, volume-averaged damage, and the displacement and damage fields to characterise expansion and damage. All comparisons used loading data for both Langer's line orientations.

## 3. Results

Figure 5 summarises the geometric, mechanical, and damage responses of all mesh geometries for Langer's lines parallel and perpendicular to the loading direction. In terms of expansion, the area ratio (Figure 5(A)) differed markedly between the geometries. At $\lambda_{\text{load}} = 1.40$, with Langer's lines parallel to loading, the area ratio decreased from HS (2.399) to IS (1.932), AS1 (1.757), AS2 (1.754), and ZM (1.414). HS therefore had an area 24.1% greater than IS and 69.7% greater than ZM. With Langer's lines perpendicular to loading, the same ordering was retained: HS (2.115), IS (1.887), AS1 (1.703), AS2 (1.647), and ZM (1.394), with HS exceeding IS by 12.1% and ZM by 51.7%. Thus, the rank ordering of area expansion was preserved between orientations, although the magnitude of expansion was reduced in the perpendicular configuration.

The Poisson's ratio results were consistent with the area ratio results and indicate the underlying deformation mechanism (Figure 5(B)). Under parallel Langer's line orientation to the loading direction, the Poisson's ratios were $-1.600$, $-0.958$, $-0.675$, $-0.670$, and $-0.029$ for HS, IS, AS1, AS2, and ZM, respectively. The magnitude for IS was 40.1% lower than that of HS, while ZM exhibited only weak transverse expansion. Under perpendicular orientation, the corresponding values were $-1.226$, $-0.887$, $-0.582$, $-0.483$, and $+0.013$, with IS 27.7% lower in magnitude than HS and ZM showing slight transverse contraction. The same pattern was evident in maximum transverse displacement: HS reached 35.07 mm and 24.73 mm in the parallel and perpendicular configurations, respectively, compared with only 0.39 mm and 0.74 mm for ZM. The larger area ratios of the auxetic geometries, particularly HS, arose from substantially greater transverse expansion during axial extension than from axial deformation alone.

The reaction force and P90 von Mises stress characterise the global and local mechanical response associated with this expansion (Figures 5(C) and 5(D)). The reaction force followed the same ordering in both orientations: IS > AS1 > AS2 > HS > ZM. Under parallel loading, IS required 144.65 N compared with 9.80 N for HS and 3.32 N for ZM; thus, HS required 93.2% less force than IS, while ZM required a further 66.2% less force than HS. Under perpendicular loading, the corresponding values were 68.51, 9.89, and 3.03 N, giving reductions of 85.6% for HS relative to IS and 69.4% for ZM relative to HS. Consequently, HS exhibited the greatest expansion with the lowest reaction force among the auxetic designs, whereas ZM's smaller expansion was associated with the lowest reaction force overall. Because the reaction force depends strongly on the specific specimen configuration, such as the number of repeating units and slit dimensions, these differences are interpreted qualitatively.

The P90 stress generally supported the same interpretation. Under parallel orientation, the values decreased from IS (3.636 MPa) to AS1 (1.213 MPa), AS2 (1.186 MPa), HS (0.725 MPa), and ZM (0.315 MPa). HS was therefore 80.1% lower than IS, while ZM was 56.6% lower than HS. Under perpendicular orientation, IS again exhibited the greatest P90 stress (1.576 MPa) and ZM the lowest (0.213 MPa); HS was intermediate at 0.950 MPa, with slightly lower values for AS2 (0.835 MPa) and AS1 (0.816 MPa). Thus, the low reaction force of HS was accompanied by a relatively low induced stress, although it did not produce the lowest P90 stress among the auxetic geometries in the perpendicular configuration.

The damage response varied more strongly with Langer's line orientation (Figure 5(E)). Under parallel orientation, maximum damage decreased from IS (0.501) to AS2 (0.182), AS1 (0.112), HS ($2.81 \times 10^{-4}$), and ZM ($4.92 \times 10^{-7}$). Maximum damage in HS was therefore approximately 99.94% lower than in IS. Under perpendicular orientation, the ranking changed to AS1 (0.793), IS (0.609), AS2 (0.0694), HS ($1.67 \times 10^{-5}$), and ZM ($7.48 \times 10^{-9}$); HS was approximately 99.998% lower than AS1. Thus, orientation substantially affected where the greatest local damage developed, particularly for AS1 and IS, whereas both HS and ZM maintained very small peak damage at the prescribed stretch.

Volume-averaged damage was compared at matched area ratios corresponding to the maximum expansion attained by ZM at a loading stretch of $\lambda_{\text{load}} = 1.40$. ZM area ratios of 1.414 and 1.394 were considered for parallel and perpendicular Langer's line orientations, respectively (Figure 5(F)). Under parallel orientation, IS exhibited the highest value ($1.20 \times 10^{-5}$), while AS2 and AS1 had values 88.8% and 94.8% lower, respectively. In the perpendicular orientation, AS1 exhibited the highest value ($7.34 \times 10^{-6}$), with AS2 and IS showing values that were 84.3% and 99.4% lower, respectively. HS and ZM exhibited near-zero volume-averaged damage in both orientations. The geometries with the highest $f_d$ also exhibited the highest maximum local damage in their respective orientations at $\lambda_{\text{load}} = 1.40$, although the $f_d$ comparison was performed at matched area ratios rather than equal stretch. The small $f_d$ values indicate limited average continuum damage across the common expansion ranges. Figure 6 also represents the maximum damage and transverse displacement for all simulated meshed skins at $\lambda_{\text{load}} = 1.40$.

The results show that expansion performance and damage depended on mesh geometry. At $\lambda_{\text{load}} = 1.40$, HS produced the largest area and the strongest auxetic response in both orientations, with area ratios 69.7% and 51.7% greater than those of ZM in the parallel and perpendicular configurations, respectively. HS also exhibited the lowest maximum damage among the auxetic geometries. ZM produced substantially less transverse expansion and the lowest area ratios, together with the lowest P90 stress and maximum damage. IS exhibited the highest P90 stress in both orientations, whereas AS1 and AS2 had intermediate values. Maximum damage was greatest in IS under parallel orientation and in AS1 under perpendicular orientation. At the same prescribed axial stretch, HS therefore combined substantial transverse expansion with comparatively low stress and damage, whereas ZM exhibited limited transverse expansion and the lowest stress and damage.

## 4. Discussion

The present study demonstrates that the large-deformation behaviour of skin meshing emerges from the interaction between mesh topology, tissue anisotropy, and damage evolution. Up to $\lambda_{\text{load}} = 1.25$, the hyperelastic and damage-based Ogden–GOH formulations showed closely comparable mechanical responses, indicating that continuum damage had limited influence on the reported responses over this loading range. At larger stretches, however, the responses became increasingly geometry- and orientation-dependent. The principal biomechanical distinction was not simply the magnitude of expansion, but how effectively each topology converted axial extension into transverse expansion and the stress and damage associated with that deformation. At the same prescribed stretch, HS produced the greatest projected expansion and strongest auxetic response, whereas ZM exhibited substantially less transverse expansion but correspondingly lower stress and damage. When compared at the same area expansion, however, ZM required substantially greater axial stretch to reach the HS area ratios, and the damage advantage of HS depended on the orientation of Langer's lines.

The geometric response provides a mechanistic explanation for these differences. At $\lambda_{\text{load}} = 1.40$, HS achieved area ratios 69.7% and 51.7% greater than ZM for Langer's lines oriented parallel and perpendicular to loading, respectively. This larger area gain was accompanied by markedly more negative Poisson's ratios and substantially greater transverse displacement, indicating that expansion arose predominantly through lateral expansion of the auxetic architecture. ZM exhibited only weak transverse expansion in the parallel configuration and slight transverse contraction in the perpendicular configuration. Consequently, its lower area ratios reflect limited lateral expansion under the prescribed axial stretch. The ZM area ratio remained below the nominal expansion ratio of 1.5:1 at $\lambda_{\text{load}} = 1.40$, indicating that this level of area expansion was not attained within the common loading range.

The force and stress responses provide further context. Among the auxetic geometries, HS combined the greatest area gain with the lowest reaction force, whereas IS required substantially greater force despite producing less expansion. ZM exhibited the lowest reaction force and P90 von Mises stress among all geometries, consistent with its limited transverse expansion. The reaction force reflects the global resistance to the imposed deformation, whereas the P90 stress characterises the upper portion of the reference-volume-weighted stress distribution. However, the reaction force depends strongly on the specific specimen configuration, such as the number of repeating units and slit dimensions, whereas the area ratio and Poisson's ratio are expected to be less sensitive to these details; therefore, force differences should be interpreted qualitatively.

The inclusion of continuum damage enabled the large-deformation response to be examined beyond the purely recoverable behaviour represented by hyperelasticity. Hyperelastic models reproduce nonlinear and anisotropic deformation but do not represent progressive damage or loss of constitutive stiffness. The present model extends the Ogden–GOH framework by associating separate damage processes with the ground-matrix and collagen-network contributions. The more rapid fibre-associated and slower matrix-associated damage identified during calibration should, however, be interpreted as constitutive processes inferred from the macroscopic mechanical response rather than direct representations of collagen rupture or matrix failure.

Relative to the present formulation, Li and Luo [16] incorporated Volokh-type functions [17,18] directly into the matrix and fibre contributions of the GOH strain energy function,

$$\Psi = \frac{\mu}{2}\left[(I_1 - 3) - \frac{(I_1-3)^{m+1}}{(m+1)(\zeta-3)^m}\right] + \frac{k_1}{k_2}\left[\exp(k_2 A^2) - 1 - \frac{2k_2 A^{n+2}}{(n+2)(\xi^2-1)^n}\right] \quad (18)$$

where

$$A = \lambda_f^2 - 1, \quad \lambda_f = \sqrt{\kappa I_1 + (1-3\kappa) I_4} \quad (19)$$

In addition to the standard GOH parameters, $m$ and $\zeta$ govern matrix damage, whereas $n$ and $\xi$ govern fibre damage, yielding nine fitted parameters in their formulation. Li and Luo reported a fitted fibre angle of $\gamma = 0.0696\,\circ$, whereas the present model prescribed the histologically reported collagen orientation of approximately $\gamma = \pm 41° \pm 5°$ for human back skin [5]. The present formulation likewise contains nine fitted parameters: five shared hyperelastic parameters and two parameters for each constituent-specific damage law. However, the staged calibration reduced the number of parameters identified simultaneously, while the collagen orientation was prescribed from independent histological measurements rather than fitted to the mechanical data. This provided closer consistency between the constitutive anisotropy and the reported collagen architecture while retaining separate descriptions of matrix- and fibre-associated damage.

The staged calibration also has limitations. Each damage law was identified from a single loading direction while previously calibrated parameters were held fixed; correlations between parameters identified in different stages were therefore not assessed, and the identified set may not be unique. Under parallel loading, the narrow fibre-damage distribution (dimensionless strain spread $\sigma_{\mathrm{GOH}} = 0.0472$) concentrated damage between $\lambda \approx 1.42$ and 1.48, producing a steep stress decline consistent with the experimental data (Figure 3C). Under perpendicular loading, however, the broader matrix-damage distribution ($\sigma_{\mathrm{Ogden}} = 0.378$) required to capture the gradual post-peak decline could not also reproduce the abrupt terminal stress drop at $\lambda \approx 1.82$. This drop occurred at nearly constant stretch and reflected macroscopic rupture of the specimen, which was not included in the matrix-damage calibration. Because the damage variables evolve continuously with local strain, the model represents progressive loss of stiffness but not crack initiation or tearing. Regions approaching full damage in the meshing simulations, particularly near slit tips, should therefore be interpreted as susceptible to failure rather than as predicted rupture.

Turning to the behaviour of the mesh geometries themselves, the geometric trends were broadly consistent with our previous hyperelastic analysis [3], in which HS also exhibited the strongest auxetic response and greatest expansion among the investigated auxetic patterns. The present study extends that observation by showing that the same geometry-dependent tendency persists when continuum damage is introduced, and the loading range is increased. Because the previous work used a different human back skin dataset, the agreement should be interpreted as evidence of structural consistency rather than direct validation. The response of HS is particularly notable because its large transverse expansion was accompanied by comparatively low stress and maximum damage among the auxetic geometries. The greater P90 stress and maximum damage of IS further demonstrate that graft coverage should be assessed alongside the stress and damage associated with expansion.

Langer's line orientation further modified these topology-dependent mechanisms. Since area ratios depend on both axial and transverse deformation, orientation-dependent changes in Poisson's ratio directly influenced the resulting graft expansion. HS remained strongly auxetic in both orientations, whereas ZM changed from weak transverse expansion to slight transverse contraction. Orientation also altered the relative damage response of AS1 and IS, indicating that local deformation and damage are sensitive to the alignment between the collagen architecture and the cut pattern. Thus, mesh geometry and material anisotropy cannot be treated as independent design variables; their interaction governs the redistribution of continuum damage during expansion assessment.

The damage measures characterised complementary aspects of continuum damage. Maximum-damage identified the greatest element-level damage, whereas volume-averaged damage characterised the overall damage across the meshed tissue, accounting for both local severity and the reference volume. A high maximum damage can coexist with a small volume-averaged value when severe damage is confined to a small proportion of the tissue. However, volume-averaged damage alone does not distinguish the spatial extent of damage from its local magnitude. Its comparison at matched area ratios therefore complements the maximum-damage comparison at equal prescribed stretch, while requiring these different deformation conditions to be considered when interpreting the results.

To compare damage at similar levels of expansion, the ZM simulations were additionally extended to $\lambda_{\text{load}} = 2.55$, aiming to match the area ratios attained by HS, which exhibited the greatest expansion at $\lambda_{\text{load}} = 1.40$ among all geometries. The orientation-specific HS area ratios were approximately 2.40 under parallel loading and 2.12 under perpendicular loading, which ZM reached only at $\lambda_{\text{load}} = 2.42$ and 2.25 (142.1% and 124.7% strain), respectively, compared with 40% strain for HS. At matched expansion, HS exhibited lower volume-averaged damage under parallel orientation but slightly higher damage under perpendicular orientation, as shown in Figure 7. The principal advantage of HS was therefore its ability to achieve comparable coverage with substantially less axial extension, accompanied by lower average damage under parallel orientation. Continued ZM loading to $\lambda_{\text{load}} = 2.55$ produced area ratios of 2.510 and 2.315, with volume-averaged damage increasing by 139.3% and 1,807.7% relative to the respective matched-area states. Although the perpendicular configuration showed a larger relative increase from its lower starting value, its final volume-averaged damage remained 87.7% below that of the parallel configuration. These results support an expansion advantage for HS in the investigated models and show that the associated damage response depends on orientation.

The comparison between ZM and the auxetic geometries must also account for differences in their initial geometric construction. The ZM slits were represented as having zero initial geometric aperture via coincident, duplicated nodes with separate connectivity, whereas the auxetic unit cells incorporated finite cutout and ligament dimensions. Sharp, zero-width slit tips would be expected to concentrate stress and promote local damage, whereas the blunter finite-width (1 mm) auxetic cuts would be expected to reduce it. The low damage in ZM, despite its sharp tips, therefore, reflects its limited deformation and stress at the same prescribed stretch, and the damage comparison between ZM and the auxetic geometries is not strictly identical. Although care was taken to keep the geometries comparable (e.g. similar slit widths, spacings and lengths, overall specimen sizes, and numbers of unit repetitions), slit length, spacing, ligament width, unit-cell dimensions, solid fraction, and specimen dimensions could not all be matched simultaneously, as keeping the slit geometry, the number of periodic repetitions and the specimen size identical across fundamentally different topologies is not feasible. A contribution of these geometric differences cannot be excluded; however, because the models were kept as similar as possible, the observed differences are expected to be governed predominantly by topology, with the effect of geometric scale likely to be minor. This consideration is particularly relevant for reaction force and total strain energy, which are structural quantities and therefore depend on the amount and distribution of material as well as on constitutive behaviour.

The present comparison used a common set of dimensions rather than dimensions optimised for each structure, and such optimisation was beyond the scope of this study. Slit length, slit spacing, cut width or aperture, ligament dimensions, and unit-cell proportions can alter the rotational and bending mechanisms responsible for auxetic expansion and can modify local stress concentration and damage development, so improvements in expansion behaviour may be achievable by tuning these parameters for each structure. A systematic parametric study would determine whether the favourable response of HS is primarily attributable to its topology or to the specific dimensions selected here, and how ZM expansion, stress, and damage change with slit dimensions. Future work could explore targeted multi-objective optimisation of each structure to maximise graft coverage while minimising internal stress and damage, rather than optimising expansion alone.

Several additional limitations should be considered. The study is computational and lacks direct experimental validation of the skin meshing geometries under matched loading and Langer's line orientations. Material parameters were calibrated from published uniaxial data for individual human back skin and therefore do not capture inter-subject variability, anatomical heterogeneity, regional thickness variations, or spatially varying collagen architecture. Nevertheless, the hyperelastic trends obtained here agreed largely with our previous analysis [3], which used material properties from a different human back skin dataset. This suggests that while local stress magnitudes are subject-specific, the large-strain pattern behaviour, such as the expansion benefit of auxetic geometries, is relatively insensitive to material properties. Skin was represented as a homogenised continuum rather than as a multilayer structure, and residual stresses, viscoelasticity, and rate dependence were not considered. Only the two limiting Langer's line orientations, parallel and perpendicular to loading, were examined; intermediate orientations may produce responses between or outside these limiting cases.

The present findings indicate that skin meshing performance should be assessed through both graft coverage (expansion ratio) and the stress and damage associated with expansion. At $\lambda_{\text{load}} = 1.40$, HS achieved the greatest area ratio and strongest auxetic response, together with the lowest maximum damage among the auxetic geometries. ZM exhibited the least expansion, the lowest P90 stress, and the maximum damage among all geometries. Extending ZM loading showed that comparable coverage required greater axial stretch than for HS, with further expansion to $\lambda_{\text{load}} = 2.55$ accompanied by increasing volume-averaged damage in both orientations. At matched coverage, HS exhibited lower volume-averaged damage under parallel orientation but slightly higher values under perpendicular orientation. HS therefore offered a favourable coverage–stretch trade for the calibrated human back skin model and selected geometries, while its damage advantage depended on orientation. Design optimisation should consider the required coverage alongside axial stretch, internal stress, and damage.

## 5. Conclusion

This study developed and applied an anisotropic damage-based Ogden–GOH formulation to investigate the large-deformation response of skin meshing while accounting for progressive continuum damage. Up to $\lambda_{\text{load}} = 1.25$, the hyperelastic and damage-based formulations produced closely comparable mechanical responses. The damage-based analyses were subsequently extended to $\lambda_{\text{load}} = 1.40$ to examine geometry- and orientation-dependent expansion, stress, and damage development.

Among the auxetic geometries at $\lambda_{\text{load}} = 1.40$, HS exhibited the greatest area ratio and strongest auxetic response among all geometries in both Langer's line orientations, while maintaining the lowest maximum damage among the auxetic geometries. ZM exhibited the lowest area ratio, P90 von Mises stress, and maximum damage among all geometries, with limited transverse expansion. To reach the area ratios attained by HS at 40% strain, ZM required 142.1% and 124.7% strain in the parallel and perpendicular orientations, respectively, and continued expansion to $\lambda_{\text{load}} = 2.55$ increased its volume-averaged damage in both orientations. At matched area ratios, HS exhibited lower volume-averaged damage under parallel orientation but slightly higher values under perpendicular orientation. HS therefore provided a favourable coverage–stretch relationship for the calibrated human back skin model and selected geometries, while its damage advantage depended on orientation. Overall, the findings highlight the interaction between mesh geometry, tissue anisotropy, and continuum damage. Skin meshing performance should be assessed through the required graft coverage and the deformation, stresses, and damage associated with achieving it. Future work should combine experimental validation, explicit fracture modelling, and systematic optimisation of slit and ligament dimensions to identify designs that increase coverage while limiting stress and damage.

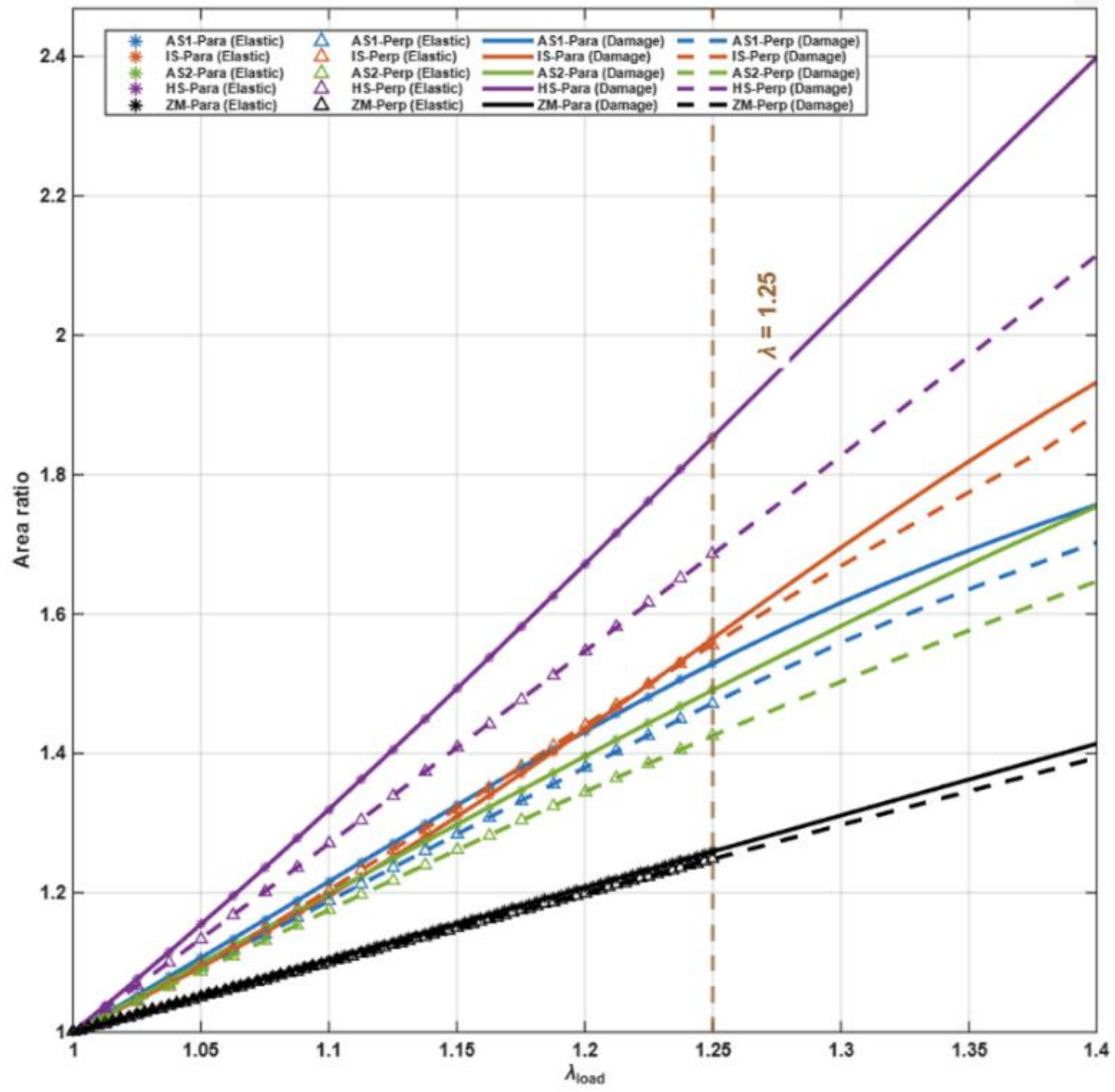
AS1-Para (Elastic)
IS-Para (Elastic)
AS2-Para (Elastic)
HS-Para (Elastic)
ZM-Para (Elastic)
AS1-Perp (Elastic)
IS-Perp (Elastic)
AS2-Perp (Elastic)
HS-Perp (Elastic)
ZM-Perp (Elastic)
AS1-Para (Damage)
IS-Para (Damage)
AS2-Para (Damage)
HS-Para (Damage)
ZM-Para (Damage)
AS1-Perp (Damage)
IS-Perp (Damage)
AS2-Perp (Damage)
HS-Perp (Damage)
ZM-Perp (Damage)
λ = 1.25
Area ratio
λload
2.4
2.2
2
1.8
1.6
1.4
1.2
1
1.05
1.1
1.15
1.2
1.25
1.3
1.35
1.4

A)

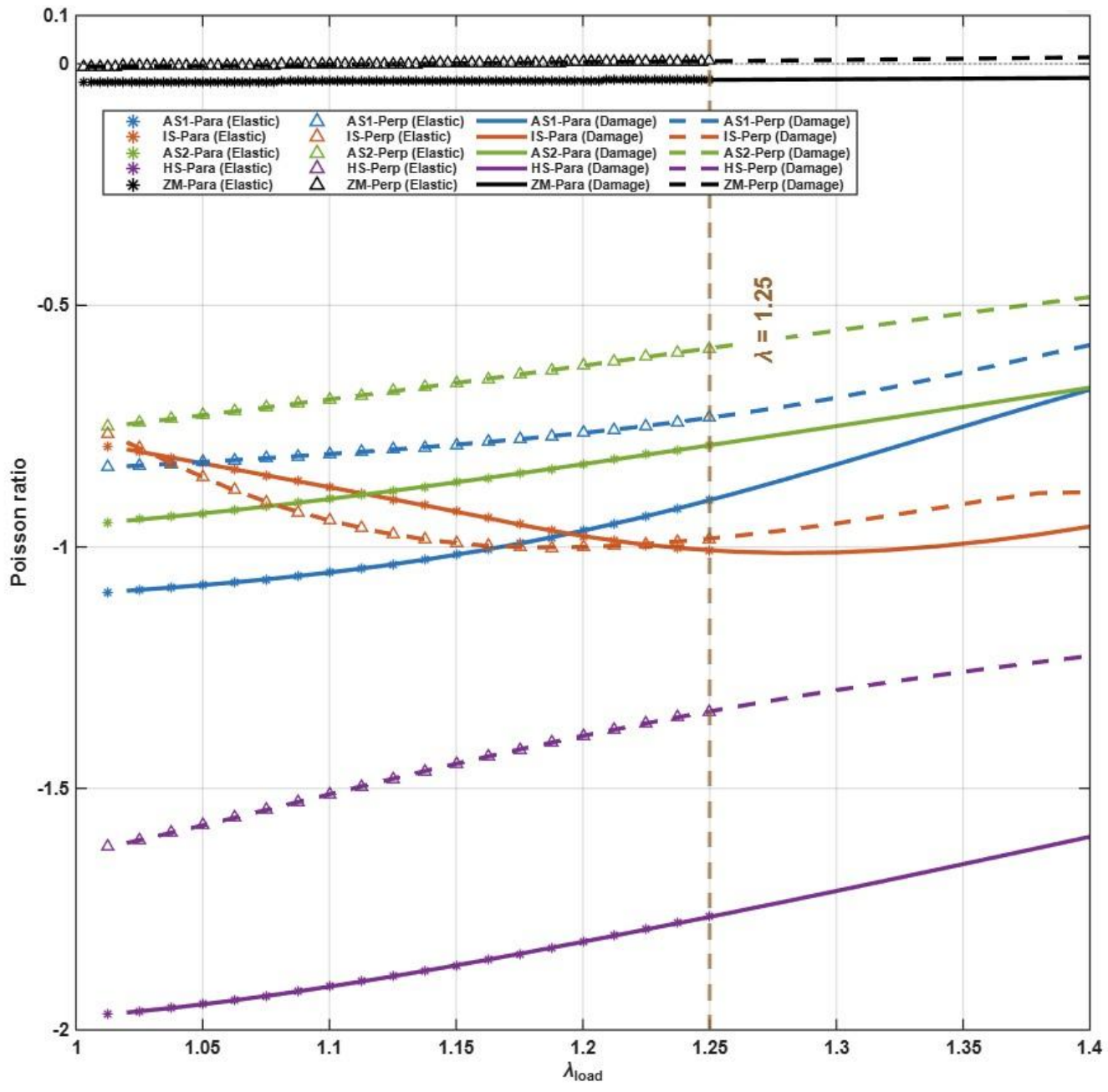
AS1-Para (Elastic)
IS-Para (Elastic)
AS2-Para (Elastic)
HS-Para (Elastic)
ZM-Para (Elastic)
AS1-Perp (Elastic)
IS-Perp (Elastic)
AS2-Perp (Elastic)
HS-Perp (Elastic)
ZM-Perp (Elastic)
AS1-Para (Damage)
IS-Para (Damage)
AS2-Para (Damage)
HS-Para (Damage)
ZM-Para (Damage)
AS1-Perp (Damage)
IS-Perp (Damage)
AS2-Perp (Damage)
HS-Perp (Damage)
ZM-Perp (Damage)
λ = 1.25
Poisson ratio
λload
0.1
0
-0.5
-1
-1.5
-2
1
1.05
1.1
1.15
1.2
1.25
1.3
1.35
1.4

B)

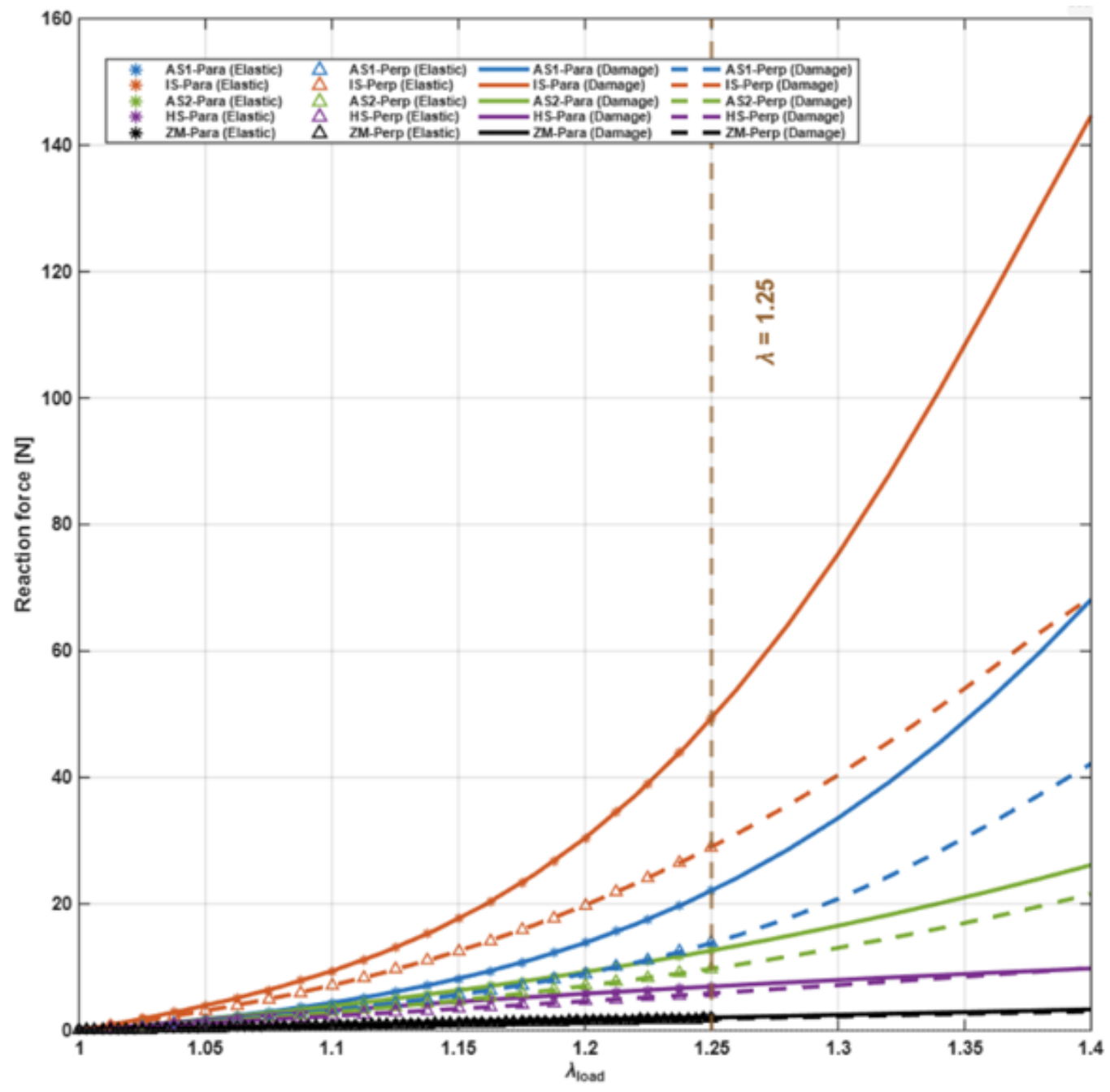

AS1-Para (Elastic)
IS-Para (Elastic)
AS2-Para (Elastic)
HS-Para (Elastic)
ZM-Para (Elastic)
AS1-Perp (Elastic)
IS-Perp (Elastic)
AS2-Perp (Elastic)
HS-Perp (Elastic)
ZM-Perp (Elastic)
AS1-Para (Damage)
IS-Para (Damage)
AS2-Para (Damage)
HS-Para (Damage)
ZM-Para (Damage)
AS1-Perp (Damage)
IS-Perp (Damage)
AS2-Perp (Damage)
HS-Perp (Damage)
ZM-Perp (Damage)
λ = 1.25
Reaction force [N]
$\lambda_{load}$
0
20
40
60
80
100
120
140
160
1
1.05
1.1
1.15
1.2
1.25
1.3
1.35
1.4

C)

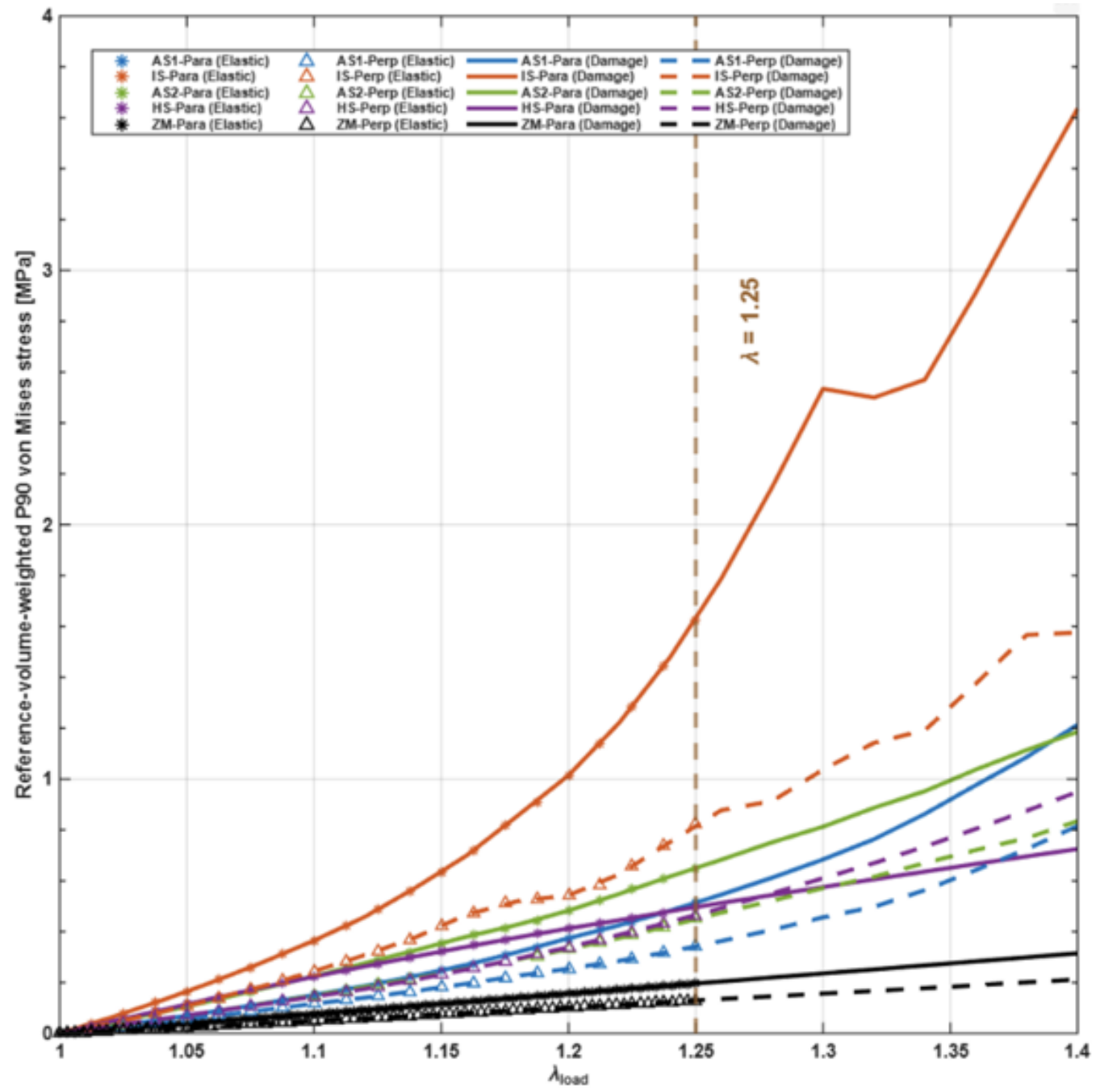

AS1-Para (Elastic)
IS-Para (Elastic)
AS2-Para (Elastic)
HS-Para (Elastic)
ZM-Para (Elastic)
AS1-Perp (Elastic)
IS-Perp (Elastic)
AS2-Perp (Elastic)
HS-Perp (Elastic)
ZM-Perp (Elastic)
AS1-Para (Damage)
IS-Para (Damage)
AS2-Para (Damage)
HS-Para (Damage)
ZM-Para (Damage)
AS1-Perp (Damage)
IS-Perp (Damage)
AS2-Perp (Damage)
HS-Perp (Damage)
ZM-Perp (Damage)
λ = 1.25
Reference-volume-weighted P90 von Mises stress [MPa]
$\lambda_{load}$
0
1
2
3
4
1
1.05
1.1
1.15
1.2
1.25
1.3
1.35
1.4

D)

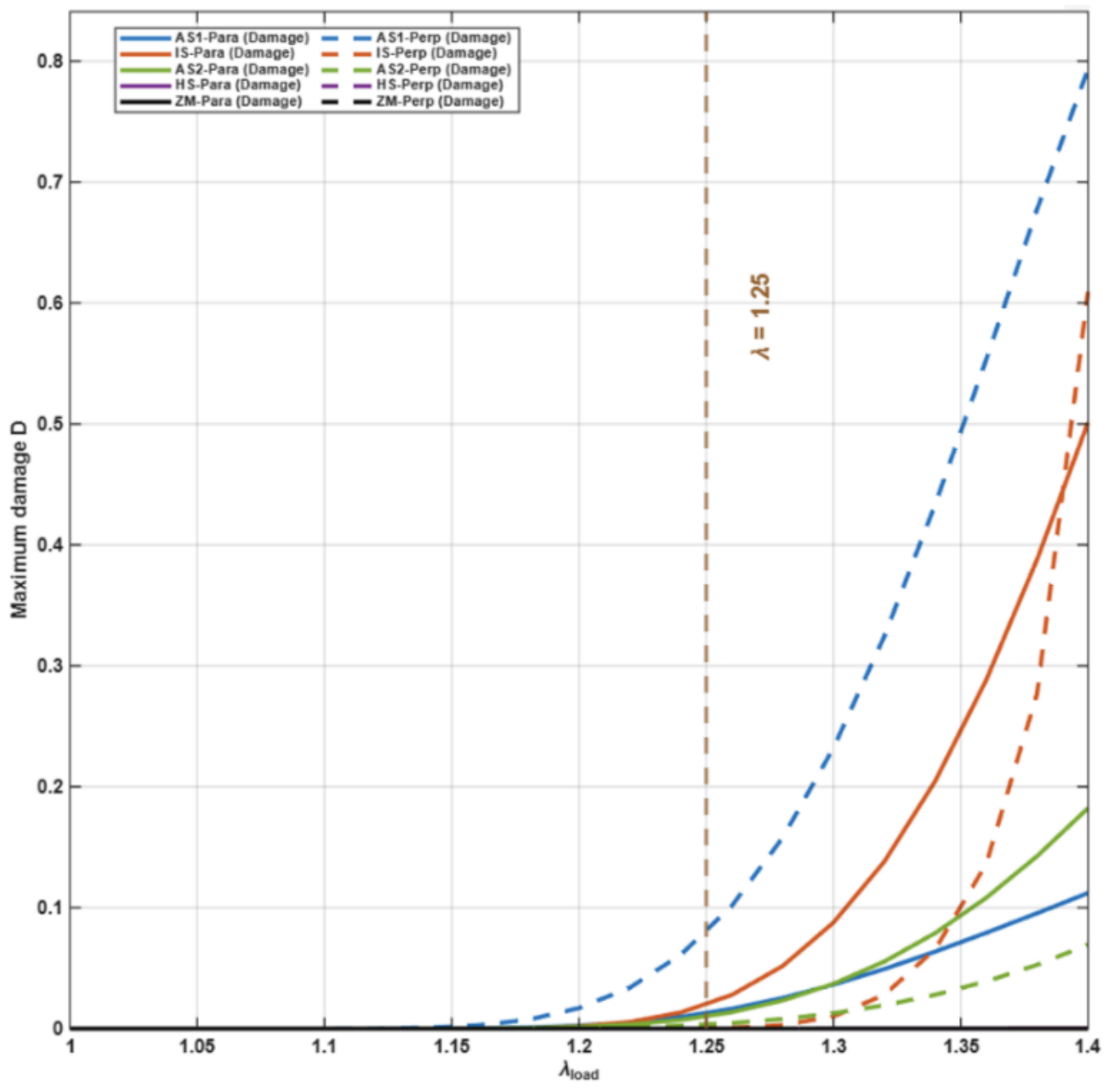

E)

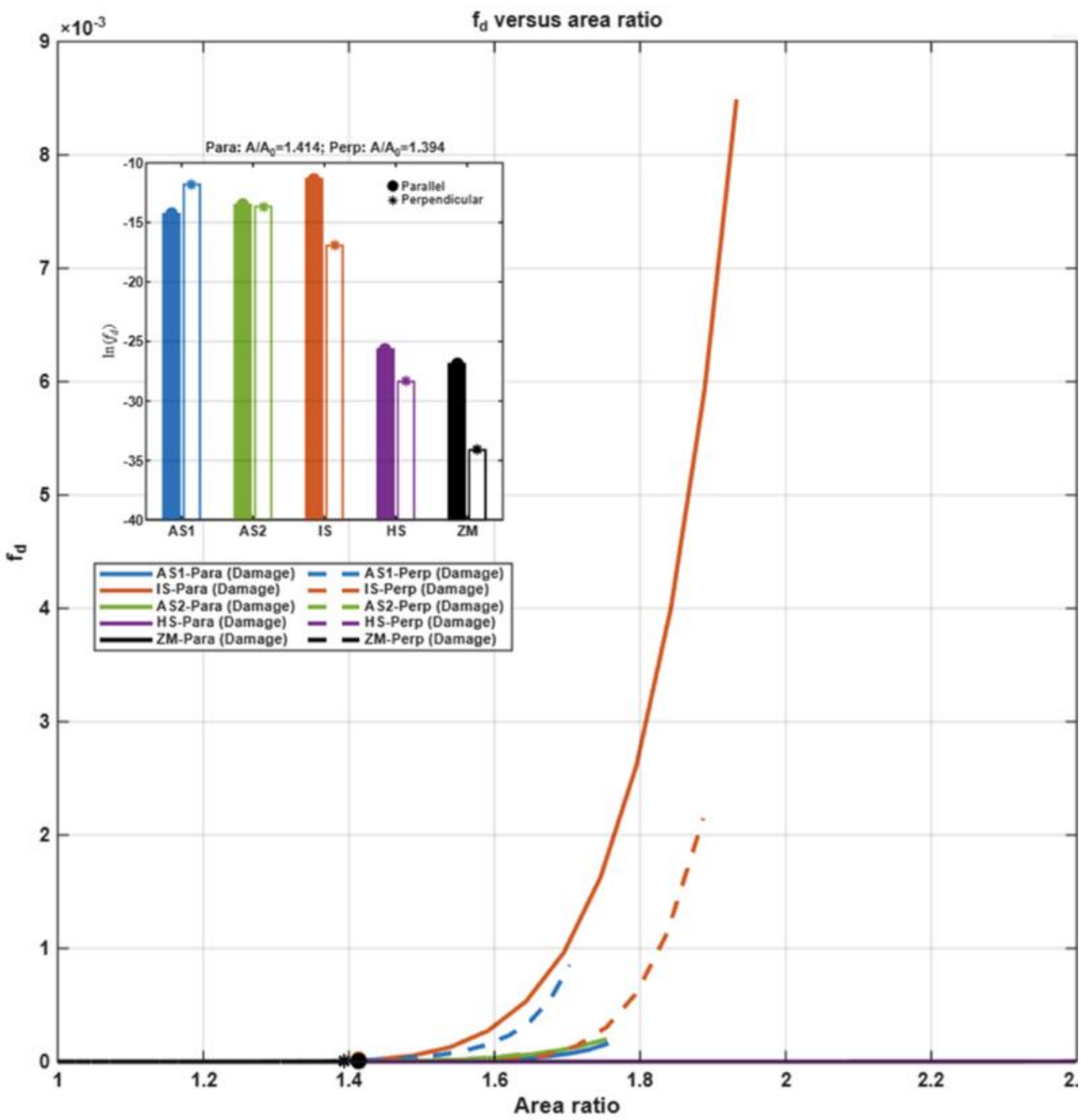

F)

**Figure 5.** Summary of the geometric, mechanical, and damage responses of AS 1st and 2nd, IS, HS, and ZM mesh geometries in the elastic range $\lambda_{\text{load}} = 1.25$ versus damaged range $\lambda_{\text{load}} = 1.40$. **(A)** Area ratio, **(B)** Poisson's ratio, **(C)** reaction force magnitude, **(D)** volume-weighted 90th percentile von Mises stress, **(E)** maximum damage, **(F)** Volume-averaged damage versus area ratio along with natural logarithm of volume-averaged damage ($\ln f_d$) at matched area ratios for parallel and perpendicular to Langer's line orientations.

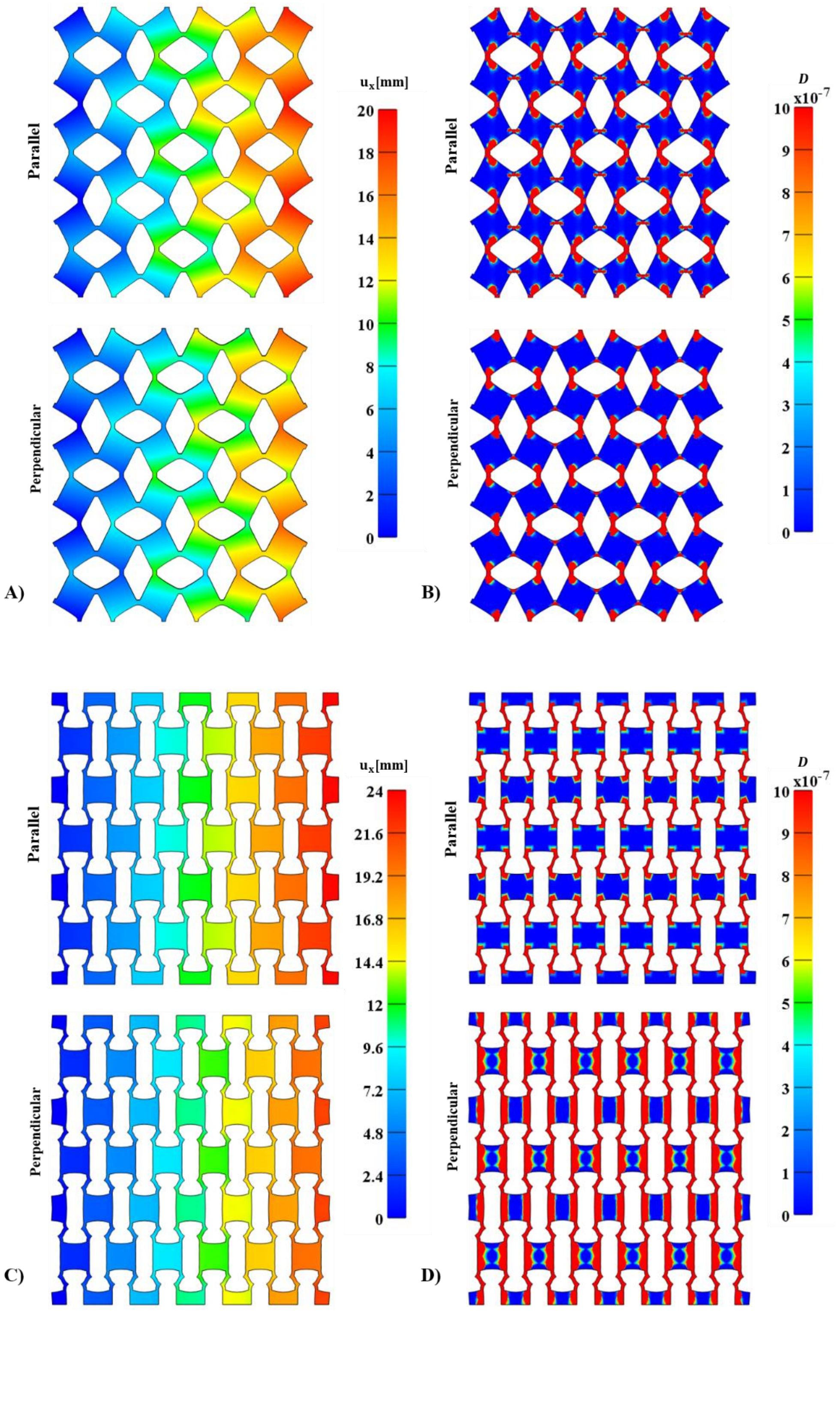
Parallel
Perpendicular
u_x[mm]
20
18
16
14
12
10
8
6
4
2
0
A)
Parallel
Perpendicular
D
x10^-7
10
9
8
7
6
5
4
3
2
1
0
B)
Parallel
Perpendicular
u_x[mm]
24
21.6
19.2
16.8
14.4
12
9.6
7.2
4.8
2.4
0
C)
Parallel
Perpendicular
D
x10^-7
10
9
8
7
6
5
4
3
2
1
0
D)

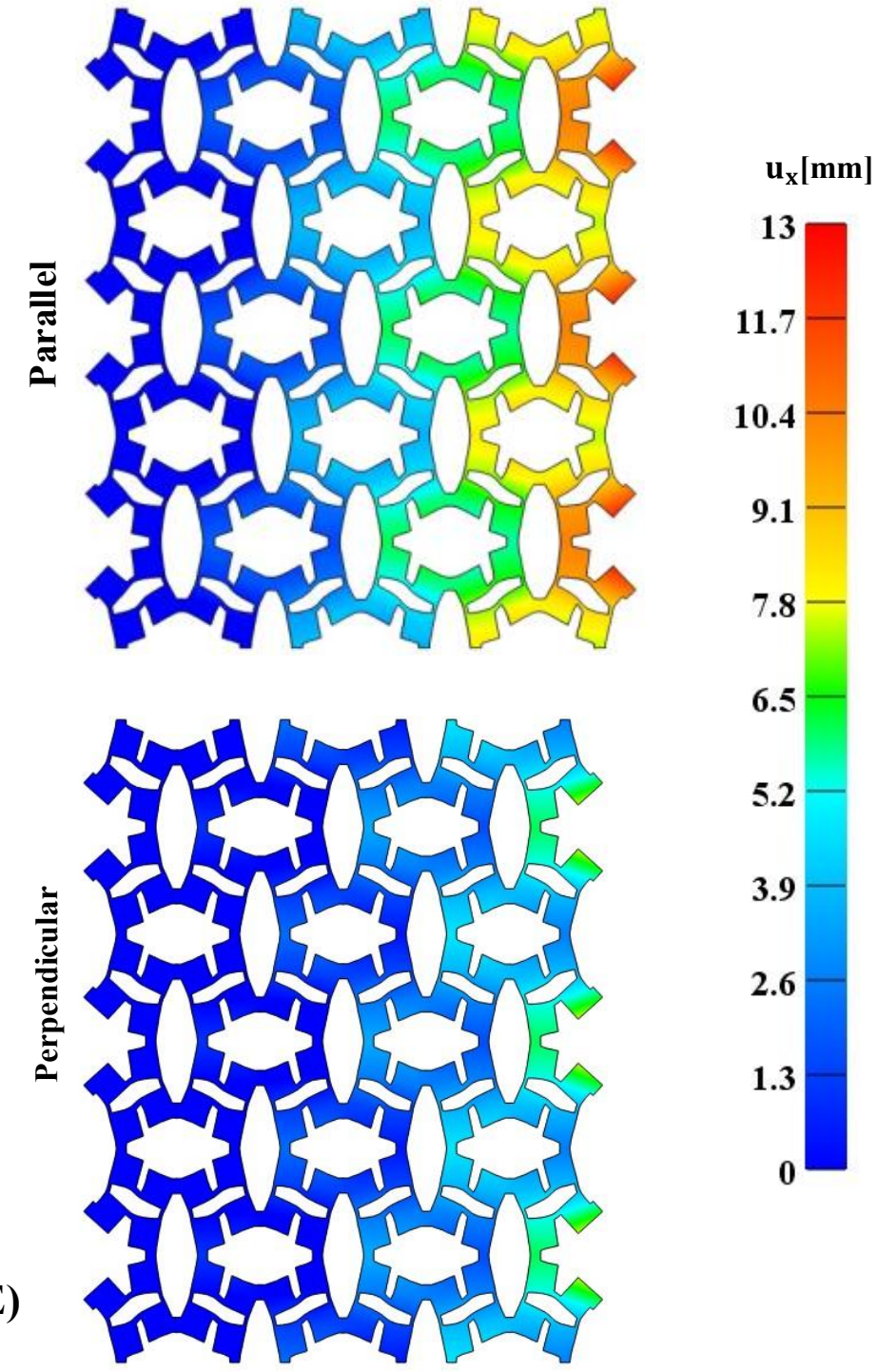
Parallel
Perpendicular
u_x[mm]
13
11.7
10.4
9.1
7.8
6.5
5.2
3.9
2.6
1.3
0

E)

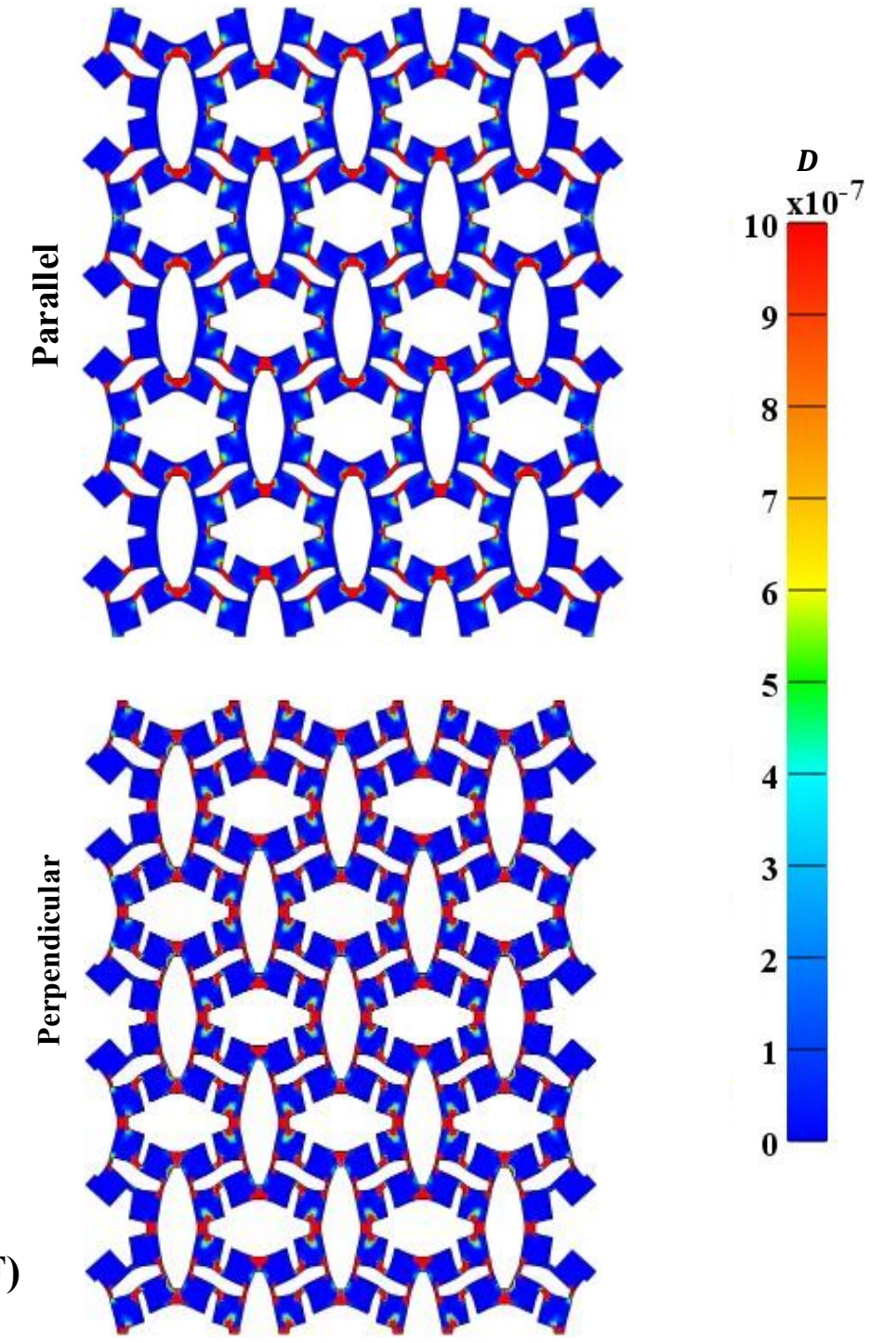
Parallel
Perpendicular
D
x10^-7
10
9
8
7
6
5
4
3
2
1
0

F)

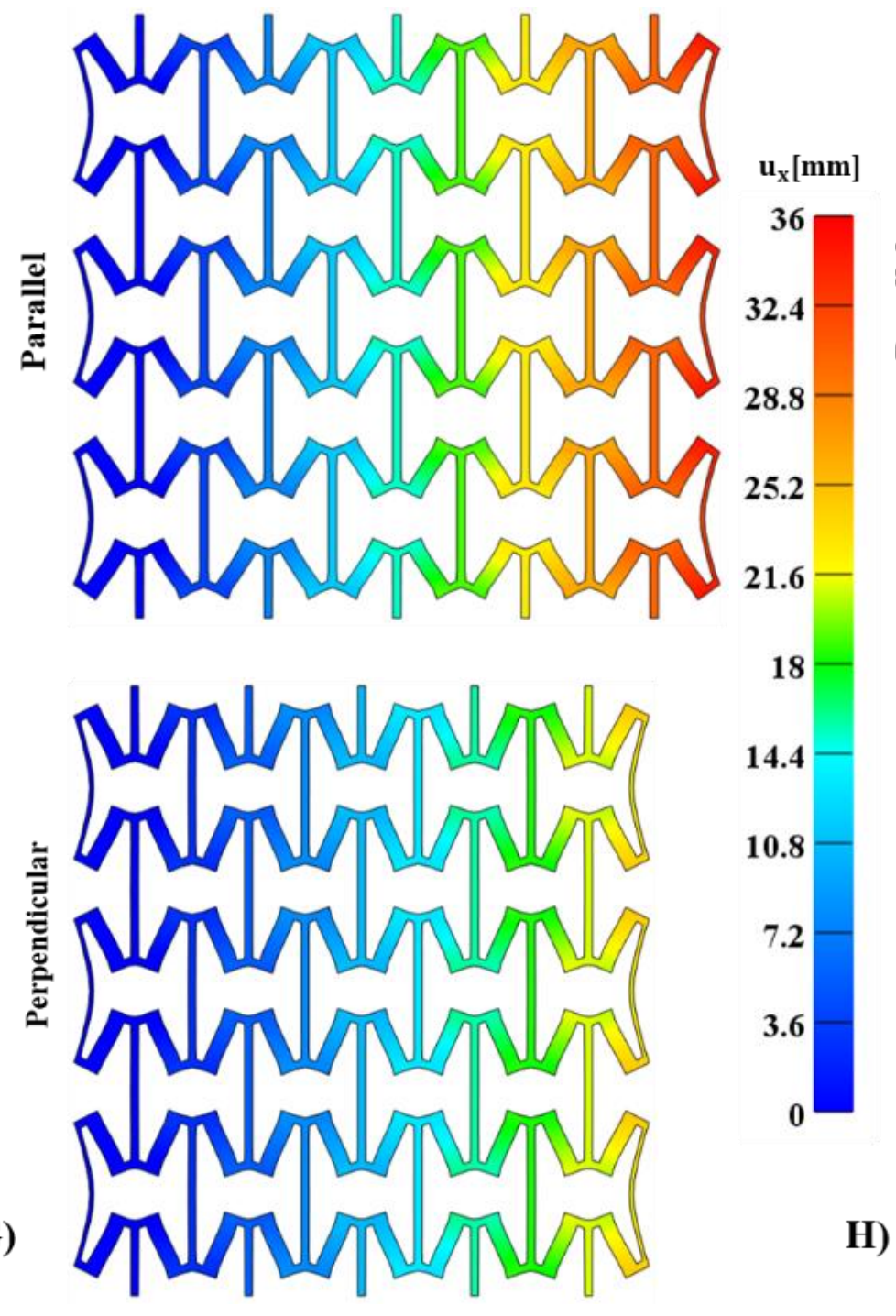
Parallel
Perpendicular
u_x[mm]
36
32.4
28.8
25.2
21.6
18
14.4
10.8
7.2
3.6
0

G)

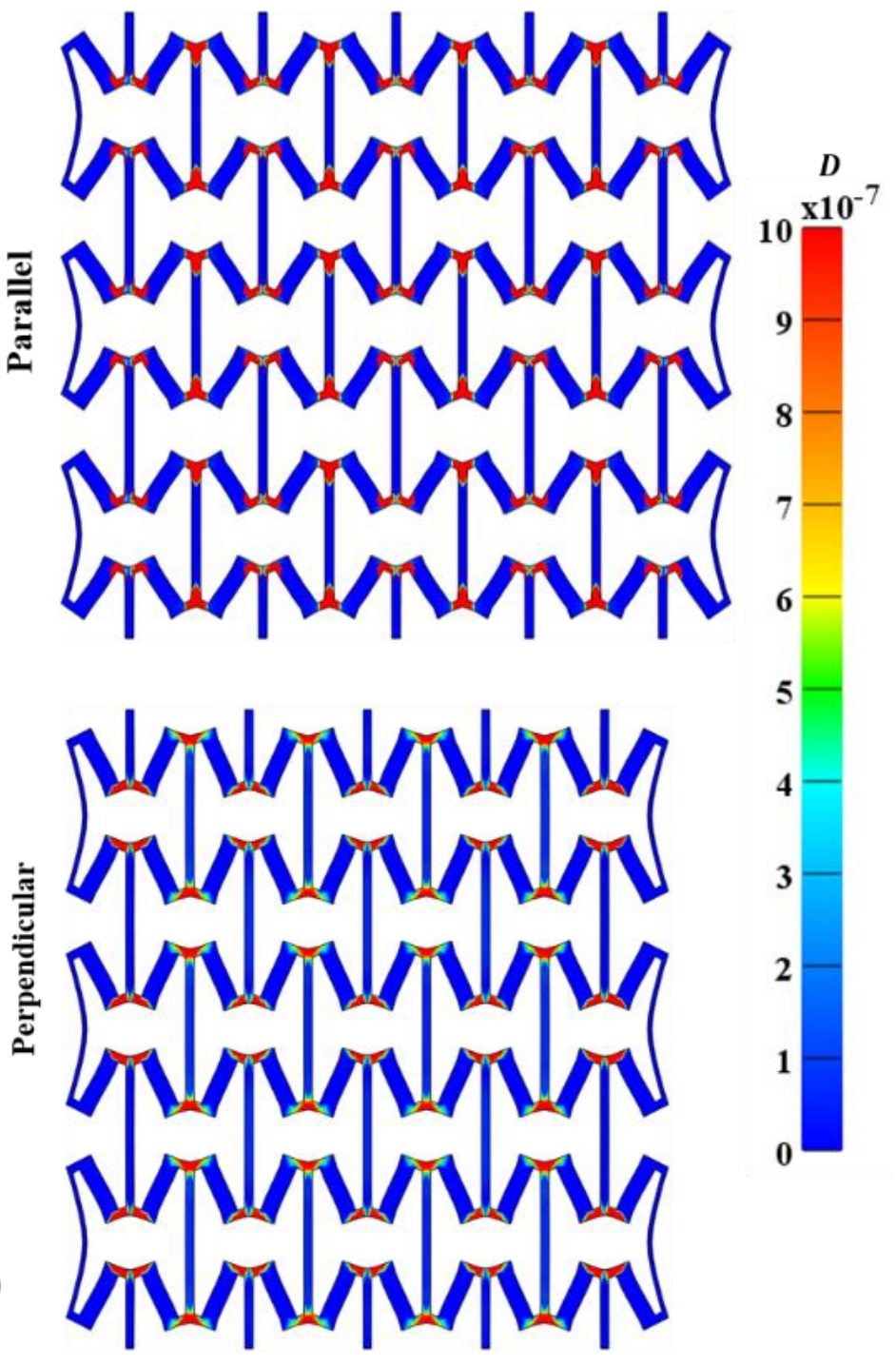
Parallel
Perpendicular
D
x10^-7
10
9
8
7
6
5
4
3
2
1
0

H)

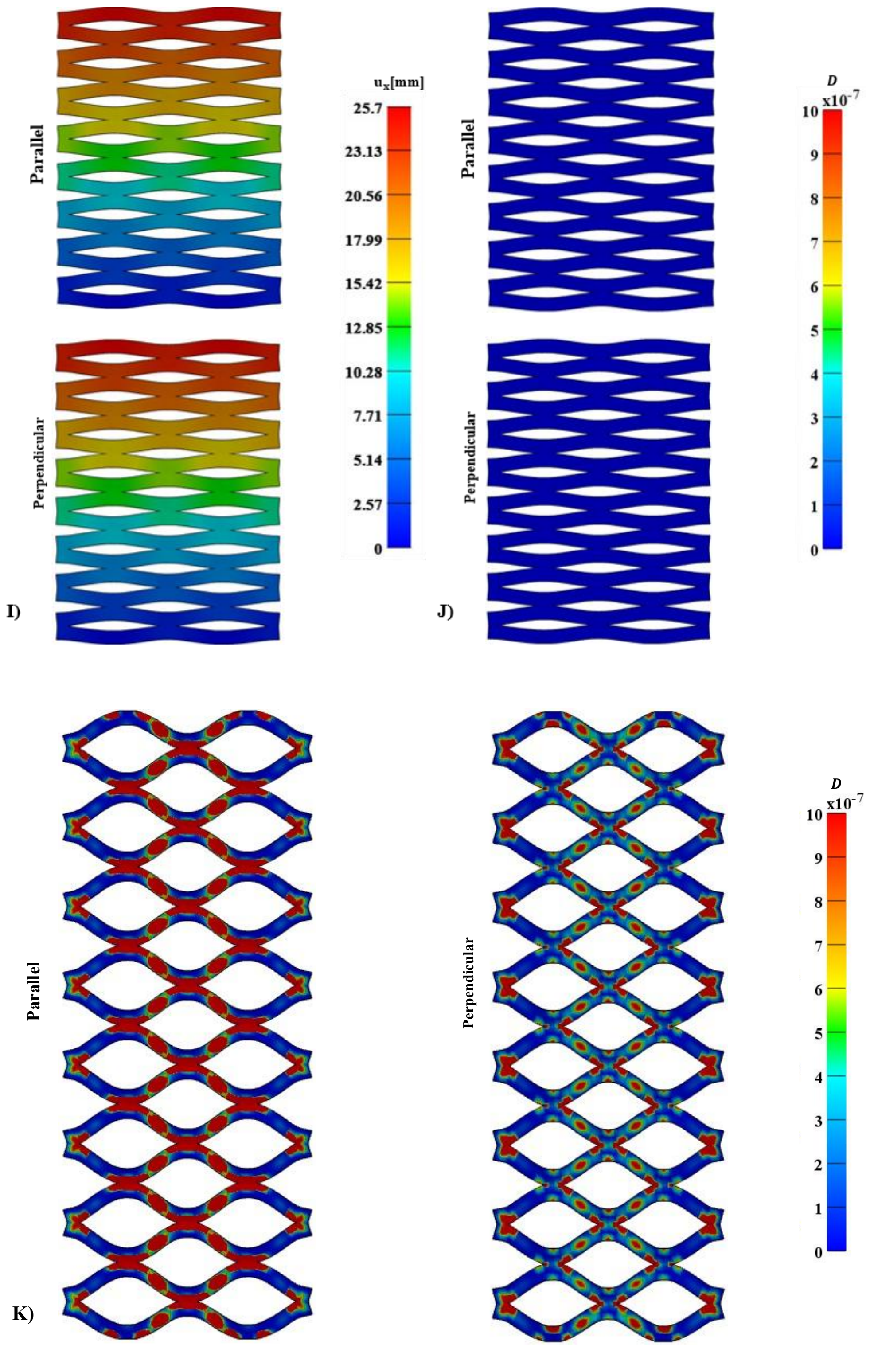

**Figure 6.** Parallel and perpendicular maximum transverse displacement (**A**, **C**, **E**, **G**, **I**) and damage distribution (**B**, **D**, **F**, **H**, **J**) of AS 1st, IS, AS 2nd, HS, and ZM at $\lambda_{\text{load}} = 1.40$. (**K**) damage distribution ZM at $\lambda_{\text{load}} = 2.55$

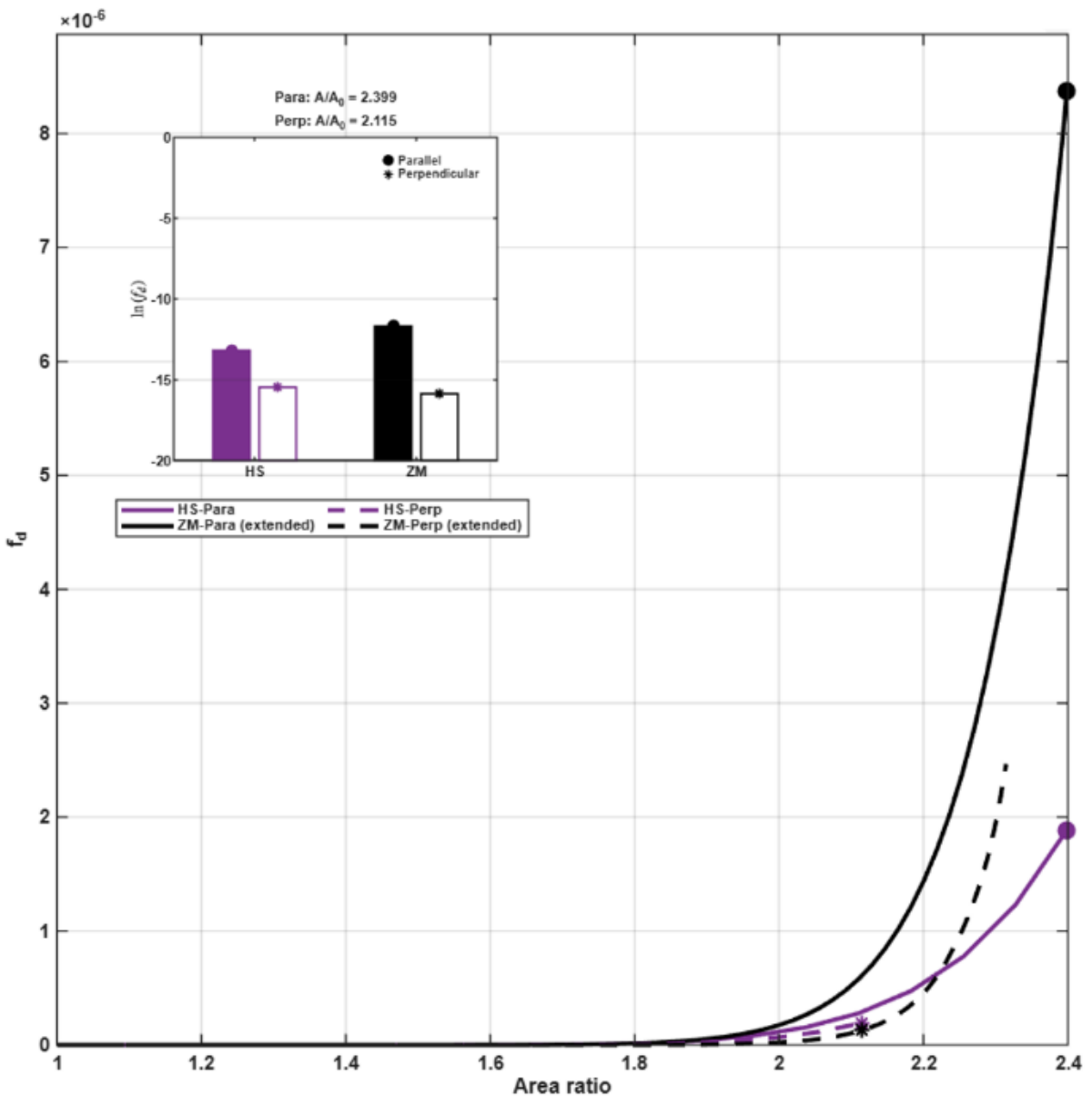

**Figure 7.** Volume-averaged damage versus area ratio for HS and extended ZM loading under parallel and perpendicular Langer's-line orientations. Circles and stars mark matched area ratios; the inset compares $\ln f_d$ at these states.

## Acknowledgments

This research was supported by the MOCHAS scholarship from Atlantic Technological University.